\RequirePackage{fix-cm}
\documentclass[twocolumn]{svjour3}          
\smartqed  
\usepackage{graphicx}
\usepackage[ T1 ]{ fontenc }
\usepackage{ amsmath, amsfonts, amssymb }
\usepackage{ stix }
\usepackage{cite}
\usepackage{ listings } 
\usepackage{ enumitem } 
\usepackage{ xspace } 
\usepackage{ xpatch } 
\usepackage{ xcolor }
\usepackage{ realboxes }
\usepackage{ bm }
\usepackage{ color }
\usepackage{ booktabs, makecell, multirow }
\definecolor{mygray}{rgb}{0.8,0.8,0.8}
\makeatletter
\xpretocmd\lstinline{\Colorbox{mygray}\bgroup\appto\lst@DeInit{\egroup}}{}{}
\makeatother

\SetSymbolFont{letters}{bold}{OML}{cmm}{b}{it}
\SetSymbolFont{operators}{bold}{OT1}{cmr}{bx}{n}

\usepackage{hyperref}
	\hypersetup{ colorlinks, 
	             citecolor = black, 
	             filecolor = black, 
	             linkcolor = black, 
	             urlcolor = black}

\allowdisplaybreaks	             
\DeclareMathOperator*{\argmin}{arg\,min}
\newcommand{\norm}[1]{\lVert#1\rVert}   
\newcommand{\lpare}[1]{\left(#1\right)} 
\newcommand{\lbra}[1]{\left[#1\right]} 
\newcommand{\dt}{\mathrm{d}t}

\newcommand{\ubold}{\mathbf{u}}

\newcommand{\fbold}{\mathbf{f}}

\newcommand{\xG}{x_{\mathrm{G}}}

\newcommand{\Psub}{_\mathrm{P}}

\newcommand{\MATLAB}{\textsc{Matlab}\xspace} 

\newcommand{\deltap}{\delta_{\mathrm{p}}}
\newcommand{\deltas}{\delta_{\mathrm{s}}}
\newcommand{\deltatildep}{\tilde{\delta}_{\mathrm{p}}}
\newcommand{\deltatildes}{\tilde{\delta}_{\mathrm{s}}}
\newcommand{\deltaph}{\delta_{\mathrm{ph}}}
\newcommand{\deltash}{\delta_{\mathrm{sh}}}
\newcommand{\xfR}{x_{\mathrm{fR}}}

\newcommand{\Bsub}{_{\mathrm{B}}}

\newcommand{\reqsub}{_{\mathrm{req}}}

\newcommand{\dsubs}{_{\mathrm{d}}}
\newcommand{\dzerosub}{_{\mathrm{d},0}}

\newcommand{\Vmtx}{\mathbf{V}}
\newcommand{\Tmtx}{\mathbf{T}}
\newcommand{\Rmtx}{\mathbf{R}}

\newcommand{\Vmtxtilde}{\tilde{\Vmtx}}

\newcommand{\ebold}{\mathbf{e}}

\newcommand{\twotwomatrix}[4]{\begin{bmatrix} \; #1 & #2 \;\; \\ \; #3 & #4 \;\;  \end{bmatrix}}
\newcommand{\threethreematrix}[9]{\begin{bmatrix} \; #1 & #2 & #3 \; \\ \;  #4 & #5 & #6 \; \\ \; #7 & #8 & #9 \; \end{bmatrix}}

\newcommand{\transp}{^{\top}}
\newcommand{\invers}{^{-1}}
\newcommand{\tworowvector}[2]{\begin{bmatrix} \; #1 \;\; #2 \;\; \end{bmatrix}\transp}
\newcommand{\threerowvector}[3]{\begin{bmatrix} \; #1 \;\; #2 \;\; #3 \;\; \end{bmatrix}\transp}
\newcommand{\fourrowvector}[4]{\begin{bmatrix} \; #1 \;\; #2 \;\; #3 \;\; #4\;\; \end{bmatrix}\transp}

\newcommand{\twovector}[2]{\begin{bmatrix} \; #1 \; \\ \; #2 \; \end{bmatrix}}
\newcommand{\threevector}[3]{\begin{bmatrix} \; #1 \; \\ \; #2 \; \\ \; #3 \; \end{bmatrix}}

\newcommand{\err}[1]{#1\dsubs}
\newcommand{\PIDcoeff}[2]{_{\mathrm{#1#2}}} 
\newcommand{\forcePID}[3]{#1\reqsub=&\,K_{\mathrm{P#1}}\err{#2}+K_{\mathrm{I#1}}\int\err{#2}\,\dt-K_{\mathrm{D#1}}#3}
\newcommand{\Vtilde}{\tilde{V}}
\newcommand{\nBtilde}{\tilde{n}\Bsub}
\newcommand{\cospsi}{\cos{\psi}}
\newcommand{\sinpsi}{\sin{\psi}}

\newcommand{\Rpsi}{\Rmtx\lpare{\psi}}

\newcommand{\Lpp}{L_{\mathrm{pp}}}

\newcommand{\csubs}{_{\mathrm{c}}}

\newcommand{\CTsub}{_{\mathrm{CT}}}

\newcommand{\XCT}{X\CTsub}
\newcommand{\YCT}{Y\CTsub}

\newcommand{\uboldeta}{\ubold_{\delta}}
\newcommand{\uboldc}{\ubold\csubs}

\newcommand{\uboldth}{\ubold_{\delta\mathrm{h}}}
\newcommand{\uboldthtilde}{\tilde{\ubold}_{\delta\mathrm{h}}}
\newcommand{\fboldct}{\fbold\CTsub}
\newcommand{\fboldreq}{\fbold\reqsub}

\newcommand{\fbolditcp}{\fbold_{\mathrm{itcp}}}

\newcommand{\Pmtx}{\mathbf{P}}

\newcommand{\nB}{n\Bsub}
\newcommand{\YB}{Y\Bsub}
\newcommand{\fboldc}{\fbold\csubs}
\newcommand{\CB}{C\Bsub}
\newcommand{\xB}{x\Bsub}
\newcommand{\Xreq}{X\reqsub}
\newcommand{\Yreq}{Y\reqsub}
\newcommand{\Nreq}{N\reqsub}
\newcommand{\Kmtx}[1]{\mathbf{K}_{\mathrm{#1}}}
\newcommand{\vbold}{\mathbf{v}}
\newcommand{\etabold}{\bm{\eta}}
\newcommand{\Wmtx}{\mathbf{W}}
\newcommand{\Zmtx}{\mathbf{Z}}
\newcommand{\etatilde}{\tilde{\etabold}}

\usepackage[pagewise, switch]{lineno}
 \newcommand*\patchAmsMathEnvironmentForLineno[1]{%
   \expandafter\let\csname old#1\expandafter\endcsname\csname #1\endcsname
   \expandafter\let\csname oldend#1\expandafter\endcsname\csname end#1\endcsname
   \renewenvironment{#1}%
      {\linenomath\csname old#1\endcsname}%
      {\csname oldend#1\endcsname\endlinenomath}}%
 \newcommand*\patchBothAmsMathEnvironmentsForLineno[1]{%
   \patchAmsMathEnvironmentForLineno{#1}%
   \patchAmsMathEnvironmentForLineno{#1*}}%
 \AtBeginDocument{%
 \patchBothAmsMathEnvironmentsForLineno{equation}%
 \patchBothAmsMathEnvironmentsForLineno{align}%
 \patchBothAmsMathEnvironmentsForLineno{flalign}%
 \patchBothAmsMathEnvironmentsForLineno{alignat}%
 \patchBothAmsMathEnvironmentsForLineno{gather}%
 \patchBothAmsMathEnvironmentsForLineno{multline}%
 }

\usepackage{ulem}

\newcommand\Erase{\bgroup\markoverwith{\textcolor{red}{\rule[.5ex]{2pt}{0.4pt}}}\ULon}

\begin{document}
\sloppy

\title{Experimental Low-speed Positioning System with VecTwin Rudder for Automatic Docking (Berthing)
}


\author{Dimas M. Rachman \and
        Yusuke Aoki \and
        Yoshiki Miyauchi \and
        Naoya Umeda \and
        Atsuo Maki
}


\institute{D. M. Rachman \and Y. Aoki \and Y. Miyauchi \and N. Umeda \and A. Maki \at
            Osaka University, 2-1 Yamadaoka, Suita \\
            Osaka 565-0871, Japan \\
            \email{dimas\_rachman@naoe.eng.osaka-u.ac.jp}  \\
}

\date{Received: date / Accepted: date}

\maketitle

\begin{abstract}
A VecTwin rudder system comprises twin fishtail rudders with reaction fins to increase its performance. With a constant propeller revolution number, the vessel can execute special low-speed maneuvers like hover, crabbing, reverse, and rotation. Such low-speed maneuvers are termed dynamic positioning (DP), and a DP vessel should be fully/overly actuated with several thrusters. This article introduces a novel and experimental VecTwin positioning system (VTPS) without making the ship fully/overly actuated. Unlike the usual dynamic positioning system (DPS), the VTPS is developed for low-speed operations in a calm harbor area. It is designed upon an assumption that the forces due to the interaction between the rudders, the propeller, and the hull are linear with the rudder angles within a range around the hover rudder angle. The linear relationship is obtained through linear regression of the results from several CFD simulations. The VTPS implements a PID controller that regulates the actuator forces to achieve the given low-speed positioning objective. It was tested in combined automatic docking and position-keeping experiments where disturbances from the environment exist. It shows promising potential for a practical application but with further improvements.

\keywords{Automatic docking \and Positioning system  \and Low-speed maneuvers \and VecTwin rudder}
\end{abstract}

\section{Introduction}
	\label{Sec:1-Intro}
    A docking (berthing) operation must be carried out at a low speed. It may involve special maneuvers like sway, rotation, and reverse. The speed is regulated by keeping the main thrust minimal, making it very difficult for a conventional underactuated vessel (fixed-pitch single-screw propeller and single rudder) to execute such special maneuvers with just the rudder. 
    
    Such special low-speed maneuvers are usually termed dynamic positioning (DP) \cite{fossen2011handbook,7835733}. They can be executed with much ease by a DP vessel: a fully/overly actuated vessel that is also equipped with multiple main thrusters and azimuth/side thrusters. The dynamic positioning system (DPS) onboard is able to maintain the vessel's \textit{pose} (position and yaw) under wave, wind, and current disturbances \cite{MIC-1980-3-1,SORENSEN1996359,TANNURI20101121,MIZUNO2018134}. It is also capable to execute a path-following operation by \textit{slowly} maneuvering the vessel. In a similar manner, a DP controller saw applications in automating docking operations \cite{MARTISEN-DOCK,BITAR202014488}. 
    
    As an alternative to the usual DPS with multiple thrusters, it is hypothesized that a VecTwin rudder system designed by Japan Hamworthy \& Co., Ltd. can be used to develop an automatic positioning system for a calm harbor area. Throughout the article, this automatic positioning system is referred to as VecTwin positioning system or VTPS. With at least one auxiliary side thruster, special maneuvers like \textit{hover}, sway (crabbing), rotation, and even reverse can be executed with a constant forward propeller revolution \cite{TankTestLSVecTwin}. This makes it suitable for automating a docking operation that has to be carried out at a low speed.
    
    Basically, the ability of the VecTwin rudder to achieve \textit{hovering} is the basis of the hypothesis. Here, the term \textit{hovering} refers to a condition when the total forces at the equilibrium are very small or equal to zero such that the ship is almost stationary. Moreover, the pair of rudder angles that results in a hovering state is termed \textit{hover rudder angle}. The standard/maker hover rudder angles are -75 degree for the port rudder and 75 degree for the starboard rudder. 
    
    The fundamental assumption that underlines the design of the positioning system is that the relationship between the rudder angle command and the resulting forces can be assumed linear, but only within a range around the hover rudder angle. In other words, the relationship is linearized around the hover rudder angle. One of the ways to obtain such a linear relationship is via multiple linear regression of the results from CFD simulations of propeller-rudder-hull interaction at a bollard pull condition (zero speed). 
    
    The procedure for designing the VTPS is quite different from that of a conventional DPS. The difference is due to the coupled nature of the VecTwin rudder that acts as a system instead of two independent rudders. It resulted in a nondiagonal force coefficient matrix and a more compact control forces allocation, which are different from those in the usual PID-based DPS \cite{fossen2011handbook,TANNURI20101121,TANNURI2006133,ALFHEIM2018116}. However, the VTPS still incorporates the traditional decoupled PID controller for each surge, sway, and yaw motion. In addition, the wave-filtering method that is important for offshore DPS is out of the scope here. This is because the VTPS is designed for a calm harbor area where oscillatory planar motions due to high-frequency waves are assumed negligible such that only the noises from the sensors exist. 
    
    Several combined automatic docking and position-keeping experiments were carried out to test the novel VTPS on a scale model equipped with a VecTwin rudder system, a fixed-pitch propeller, and a bow thruster. These experiments were carried out in the Osaka University experiment pond (Inukai pond) where the VTPS saw action under random disturbances from the environment, dominantly from the wind and gusts. Based on its performance during the experiments, the VTPS can be a potential addition to other automatic docking experimental schemes \cite{MARTISEN-DOCK,ahmed2014experiment,MIZUNO2015305,Sawada2021} and various numerical schemes \cite{shouji1992automatic,IM2018235,MO-NMPC,WANG2022111021}.
    
    Overall, this article serves as a preliminary investigation to test the hypothesis directly in a scale model experiment. On this basis, the VTPS was tested first to see that it indeed practically works well, which is why the term \textit{experimental} is used to indicate that the VTPS is, in some sense, a working prototype. It is directly verified and validated in the experiments, bypassing verification with simulation. One may think of it as reverse engineering rather than the conventional design steps. 
    
    \begin{figure}[ht!]
    \centering
    \includegraphics[width=0.95\columnwidth]{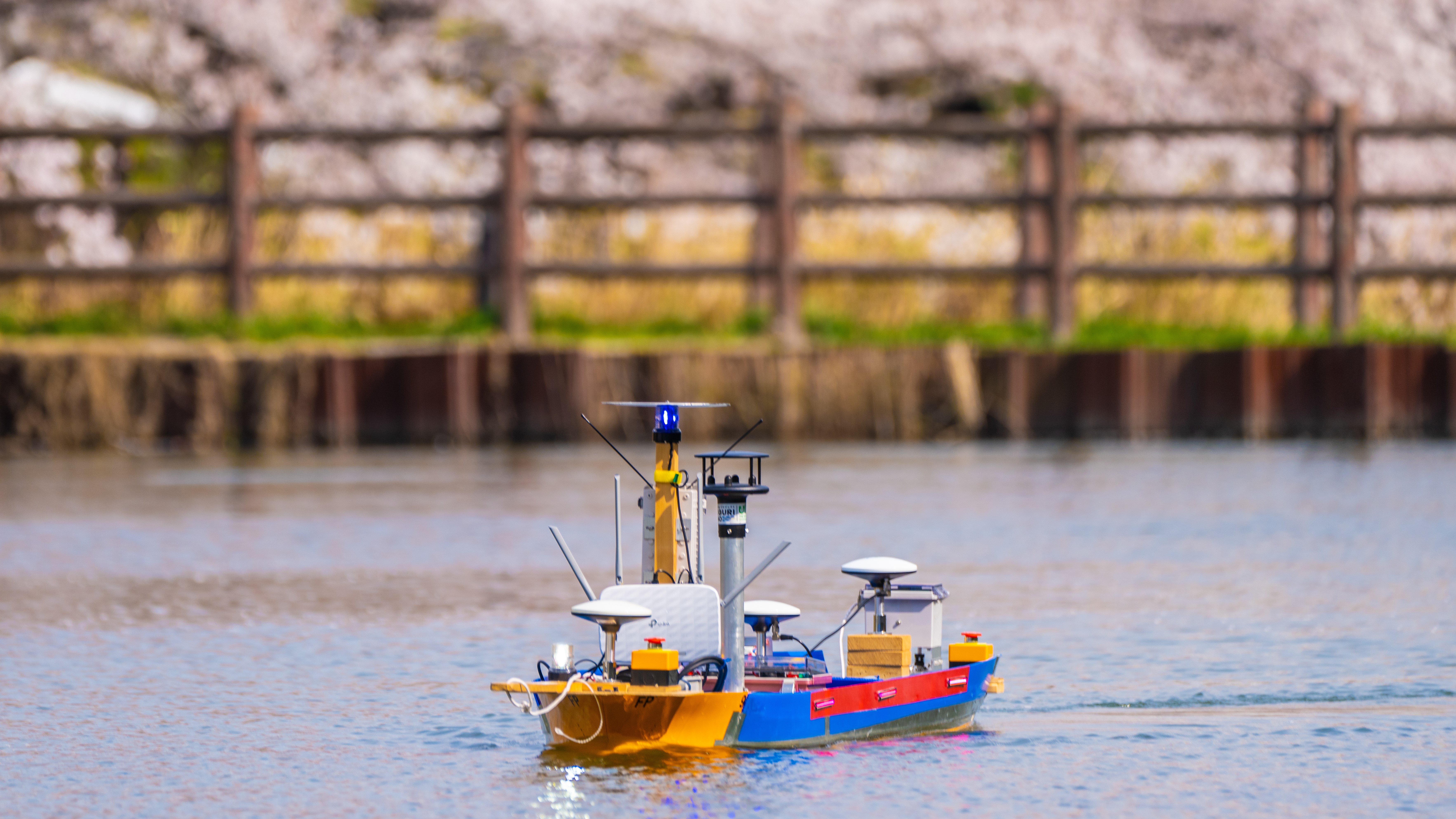}
    \caption{1:50 scale model ship that is equipped with a VecTwin rudder system, one fixed-pitch propeller, and one bow thruster.}
    \label{fig:TakaokiSakura}
    \end{figure}
    
    \begin{table}[ht!]
        \caption{Particulars of the 1:50 scale model.}\label{tab:A1-TakaokiPP}
		\begin{tabular*}{\columnwidth}{ @{} l l @{} }
			\hline\noalign{\smallskip}
			 Parameters & \\
			 \noalign{\smallskip}\hline\noalign{\smallskip}
			Length between perpendiculars: $L_{\mathrm{pp}}$ (m) & 3.000  \\
			Breadth (m) & 0.484  \\
			Draft (m) & 0.172 \\
			Longitudinal center of gravity from midship: $\xG$ (m) & 0.094 \\
			Longitudinal position of rudders from $\xG$: $\xfR$ (m) & -1.657 \\
			Longitudinal position of bow thruster from $\xG$: $\xB$ (m) &  1.263 \\
			\noalign{\smallskip}\hline
		\end{tabular*}
    \end{table} 
    
\section{Design Overview}
	\label{Sec:2-DesignOverview}
    The preliminary design of the experimental VTPS is intended for a 3 m scale model of a ship with a VecTwin rudder system and a bow thruster (Fig. \ref{fig:TakaokiSakura}). Its particulars are shown in Table \ref{tab:A1-TakaokiPP}. The VTPS consists of the following four basic modules:
    \begin{itemize}
        \item Full-state measurement module: composed of multiple high-precision global navigation satellite system (GNSS) that measure the position and velocities, and a fiberoptic gyroscope that measures the yaw and yaw rate. This module is equipped with selectable linear filters to suppress the noise from the sensors to some extent.
        \item Path-following reference (setpoint) module: serves to decide which next setpoint (position and yaw) should be selected from a database of paths/waypoints. This will be covered in section \ref{Sec:6-PathRef}.
        \item Controller module: traditional decoupled PID controller that regulates the required forces to achieve the control objective; each for surge, sway, and yaw motions. This will be covered in section \ref{Sec:5-PIDControl}.
        \item Control forces allocation module: serves to distribute the required forces to the available actuators and maps them into the corresponding actuator commands (rudder angle and bow thruster revolution number). This will be covered in section \ref{Sec:3-CommandForcesRel} and section \ref{Sec:4-ForcesAlloc}.
    \end{itemize}
    
    All modules are integrated into a robotic operating system (ROS) and \MATLAB Simulink environment. Note that the propeller is kept constant in forward mode such that it is not available as a free actuator. This is one of the highlighted points in the design of the VTPS that is different from the usual DPS. 

\section{Command-Forces Relationship}
    \label{Sec:3-CommandForcesRel}
    This section explains the relationship between the actuator commands (rudder angle and bow thruster revolution number) and the resulting forces. Let us now define the actuator commands as
    \begin{align}\label{eq:actuatorcmdvect}
        \uboldc=\threerowvector{\deltap}{\deltas}{\nB}, 
    \end{align}
    where $\deltap$ and $\deltas$ denote the port and starboard rudder angle (degree), respectively, and $\nB$ denotes the bow thruster revolution number (rps). It is also convenient to vectorize a pair of rudder angles as
    \begin{align}\label{eq:deltapairvect}
        \uboldeta=\tworowvector{\deltap}{\deltas}.
    \end{align}
    
    Actually, the propeller is also an actuator. However, the propeller revolution number, denoted by $n$, is kept constant and unchanged in forward mode. This renders it unavailable as a free actuator from the point of view of control. Nevertheless, the interaction between the dynamics of the propeller with the rudder and the hull is taken into account. The later passages will explain more about this interaction.
    
    \subsection{VecTwin Rudder}
		\label{SubSec:3.1-VecTwinRudder}
		VecTwin rudder system is a high-performance rudder system that comprises twin fishtail rudders with reaction fins attached \cite{JAPANHAM}. Each rudder can rotate independently up to 140 degree: $\deltap\in\lbra{\;35,\;-105\;}$ and $\deltas\in\lbra{\;-105,\;35\;}$. This wide range of rudder, together with the reaction fins, greatly increases the maneuverability of the ship \cite{largevesselVecTwin}. 
		
		Complex interactions between the forces exerted by the rudders, the propeller, and the hull make it possible for the ship to hover and reverse at a constant forward propeller revolution. However, the complexity makes it quite challenging to model the relationship between the given rudder angle command and the resulting forces theoretically \cite{largevesselVecTwin}. Alternatively, the command-forces relationship for the VecTwin rudder can be obtained empirically via multiple linear regression of the results from steady-state CFD simulations.

        Tank tests to obtain this relationship can be conducted as shown \cite{TankTestLSVecTwin}, which is indeed preferable. However, when the resources and opportunity to conduct such tests are unavailable, one can opt to estimate the relationship via CFD simulations; under the condition that they are simulated with acceptable fidelity and its implementation results in the desired outcomes.

		\subsubsection{CFD Simulations}
			\label{SubSubSection:3.1.1-CFDSim}
			Based on the fundamental assumption, the CFD simulations should be executed for combinations of rudder angles at a range around the standard hover angle ($\deltap=\;$-75 degree and $\deltas=\;$75 degree). In this regard, the simulations were performed with combinations of $\deltap\in\lbra{\;-80,\;-70\;}$ and $\deltas\in\lbra{\;70,\;80\;}$ at 5-degree intervals. Meanwhile, the propeller was constant at $n=\;$ 10 rps. In total, there were nine combinations of rudder angles, thus nine CFD simulations.
    
			\begin{figure}[htbp]
				\centering
				\includegraphics[width=\columnwidth]{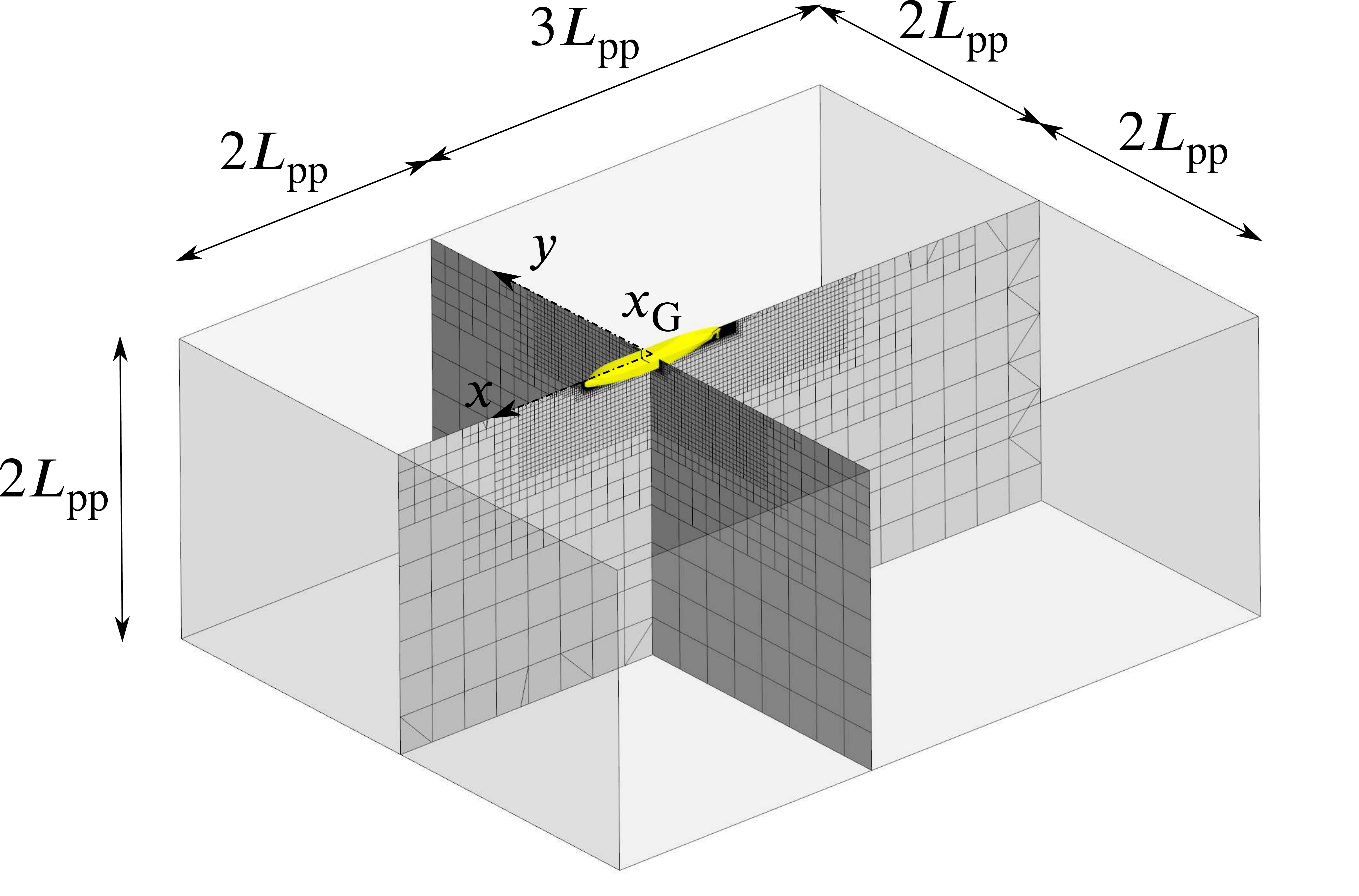}
				\caption{Size of the CFD computational domain.} 
				\label{fig:domain}
			\end{figure}
			
			A commercial solver STAR-CCM+ ver.2021.2.3 was used to perform the simulations with steady Reynolds-averaged Navier-Stokes (RANS) approach. The computational domain was rectangular with the ship's center of gravity $\xG$ as the origin. The domain size was: $-3\Lpp\geq \xG\geq 2\Lpp$ in longitudinal direction; $-2\Lpp\geq \xG \geq 2\Lpp$ in lateral direction; and 2$\Lpp$ in vertical direction, where $\Lpp$ is the length of the ship. Fig. \ref{fig:domain} visualizes this computational domain.

			The boundary conditions and the initial conditions of the computational domain were set such that the ship is at a bollard pull condition, i.e., the free-stream velocity $U_{\infty}$ and the ship's speed $u$ are both zero $\lpare{U_{\infty}=u=0}$. No-slip wall condition was applied on the surface of the hull. Pressure boundary conditions were applied to the sides of the domain and symmetric boundary conditions were applied to the top and bottom of the domain. The mesh density was verified and validated with captive model tests shown in \cite{AokiVV}. In total there were 11 million cells that built up the computational domain. For more details of the settings, we referred to the simulation setup and conditions presented in \cite{AokiVV} on free-running CFD simulations that have been verified and validated with the captive model tests of the same ship.
			
			\begin{figure}[htbp]
				\centering
				\includegraphics[width=\columnwidth]{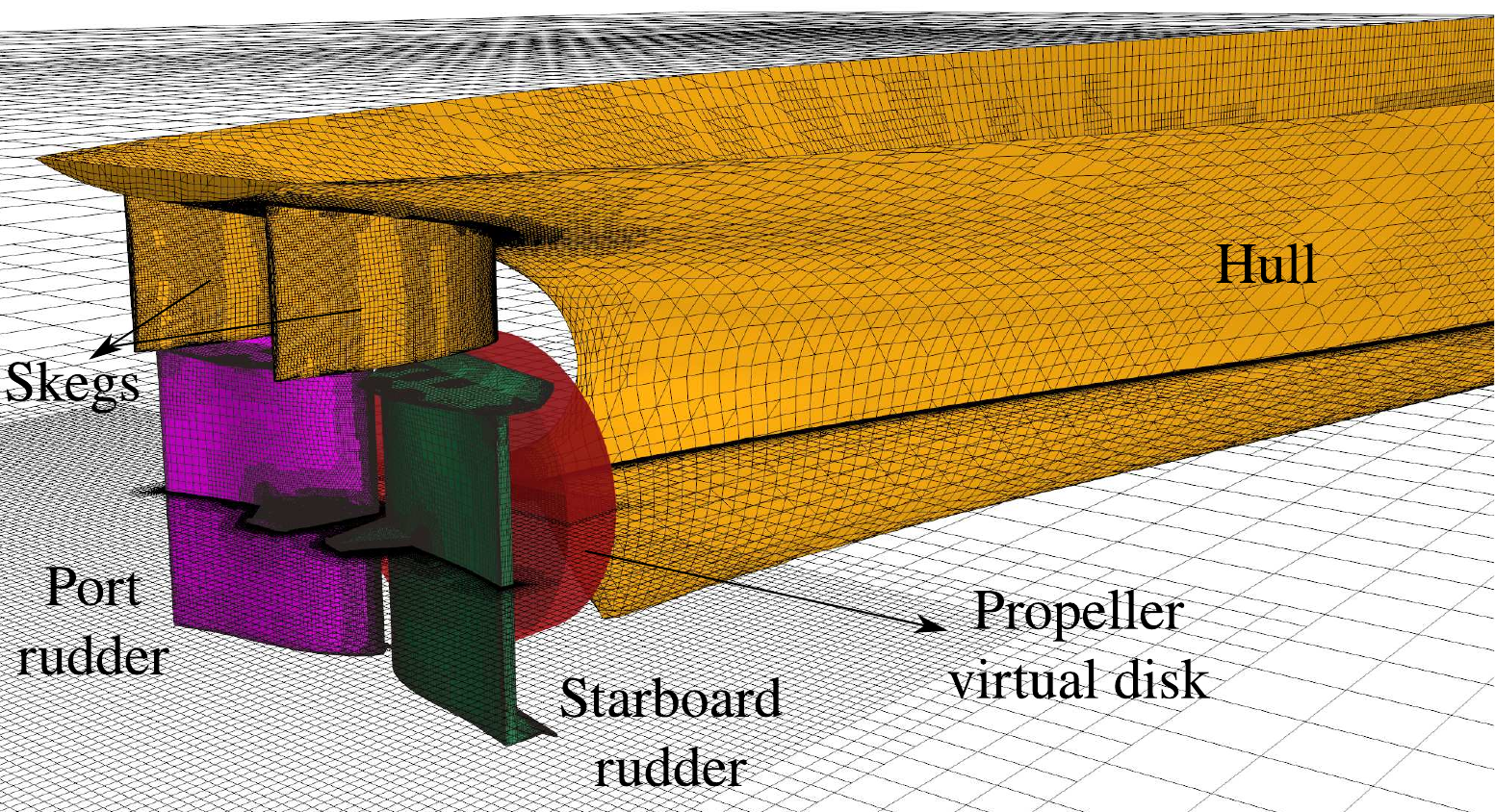}
				\caption{Hull, rudders, skegs, and propeller in the CFD simulation.} 
				\label{fig:sternview}
			\end{figure}
			
			In addition, the propeller was modeled as a virtual disk (see Fig. \ref{fig:sternview}). Its induced forces were modeled using blade element method \cite{Rajagopalan1993three} to capture the fluid interaction between the propeller, rudders, and hull accurately within moderate computational cost. The arising turbulence was modeled using shear stress transport (SST) $k$-$\omega$ model \cite{Menter2003}. Moreover, the simulations were done under double-body assumption such that the free-surface effect is negligible.

			\begin{figure}[htbp]
				\centering
				\includegraphics[width=0.53\columnwidth]{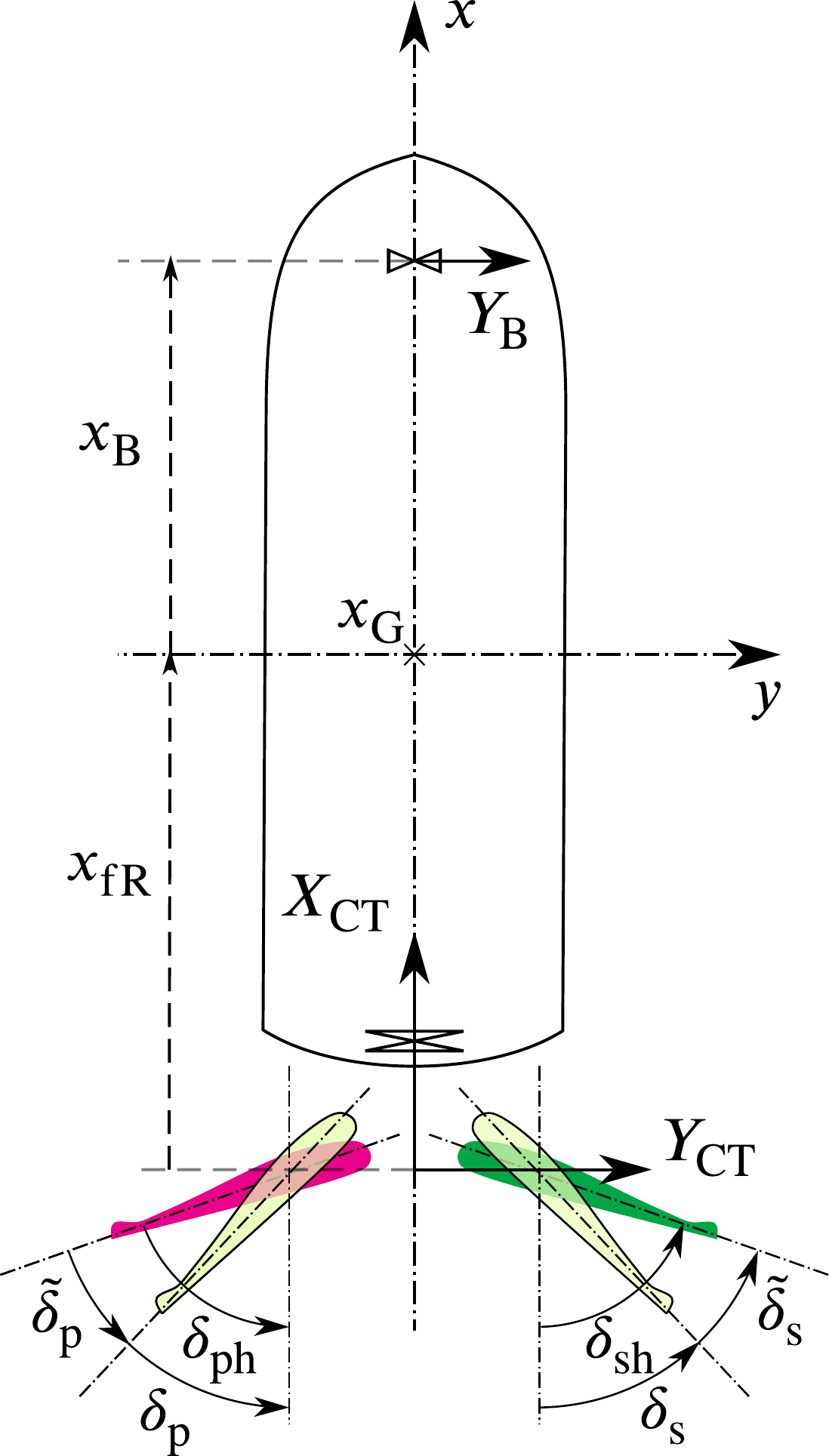}
				\caption{Ship-fixed coordinate system and rudder angle definition. } 
				\label{fig:coordinatesystem}
			\end{figure} 
			
			Suppose a ship-fixed coordinate system is introduced with the ship's center of gravity $\xG$ as the origin (Fig. \ref{fig:coordinatesystem}). In this coordinate system, the total forces due to the propeller-rudder-hull interaction from the CFD simulations, denoted by $\fboldct$, are defined as follows,
			\begin{align}\label{eq:CFDforcevect}
				\fboldct = \tworowvector{\XCT}{\YCT},
			\end{align}
			where $\XCT$ and $\YCT$ denote the total forces in surge and sway direction, respectively. These forces are assumed to act at a position $\xfR$ from $\xG$ along the centerline, which is the longitudinal position of the rudders from $\xG$ (see Table \ref{tab:A1-TakaokiPP}). The forces from all nine CFD simulations are summarized in Table \ref{tab:CFDSummary}.

			\begin{table}[h]
				\caption{Summary of the total forces from the CFD simulations at $n=\;$10 rps.}\label{tab:CFDSummary}
				\begin{tabular*}{\columnwidth}{ @{\extracolsep{\fill}} l l l l }
					\hline\noalign{\smallskip}
					$\deltap$ (deg) & $\deltas$ (deg)  & $\XCT$ (N) & $\YCT$ (N)          \\
					\noalign{\smallskip}\hline\noalign{\smallskip}
					-80 & 70 & 0.0735  & -0.0400 \\
					-80 & 75 & -0.0620 & -0.0482 \\
					-80 & 80 & -0.2061 & 0.0507  \\
					-75 & 70 & 0.1298  & 0.0544  \\
					-75 & 75 & -0.0063 & 0.0655  \\
					-75 & 80 & -0.1089 & 0.1816  \\
					-70 & 70 & 0.3724  & 0.1152  \\
					-70 & 75 & 0.1380  & 0.1552  \\
					-70 & 80 & 0.0030  & 0.2678 \\
					\noalign{\smallskip}\hline
				\end{tabular*}
			\end{table}    

		\subsubsection{Linearity Around Hover Rudder Angle}
			\label{SubSubSection:3.1.2-LinearHVR}
			Under the assumption that $\XCT$ and $\YCT$ are uncorrelated to each other, one can construct the following system of linear regression models,
			\begin{align}\label{eq:regressionsystem}
				\fboldct = \Vmtxtilde \uboldeta + \fbolditcp,
			\end{align}
			where $\fbolditcp$ is the intercept of the model (the total forces when $\uboldeta=\bm{0}$). Here, $\Vmtxtilde$ is a matrix whose rows are the regression coefficients obtained from multiple linear regression of the data points in Table \ref{tab:CFDSummary}. The entries of $\Vmtxtilde$ are as follows,
			\begin{align}\label{eq:vmtxtilde}
				\Vmtxtilde=\twotwomatrix{\Vtilde_{11}}{\Vtilde_{12}}{\Vtilde_{21}}{\Vtilde_{22}} = \twotwomatrix{0.0236}{-0.0296}{0.0192}{\hphantom{-}0.0123}.
			\end{align}
			
			As per definition of hover rudder angle, one can also obtain a corrected hover rudder angle from (\ref{eq:regressionsystem}) by setting all the forces to zero, i.e., $\fboldct=\bm{0}$. This gives us
			\begin{align}\label{eq:hoverangleCFD}
				\uboldth = -\Vmtxtilde\invers\fbolditcp,
			\end{align}
			where $\uboldth$ is the hover rudder angle. Let $\deltaph$ and $\deltash$ be the port and starboard rudder angle, respectively, that result in a hovering state. With this, $\uboldth$ can be defined as
			\begin{align}\label{eq:hoverangle}
				\uboldth=\tworowvector{\deltaph}{\deltash},
			\end{align}
			which, upon solving (\ref{eq:hoverangleCFD}), is equal to
			\begin{align}\label{eq:hoveranglevalue} 
				\uboldth = \tworowvector{-78.60}{73.38}.
			\end{align}
			Notice that the hover rudder angle obtained from the regression is almost the same as the standard hover angle $\tworowvector{-75}{75}$.
			
			The intercept $\fbolditcp$ has to be excluded since we are only interested in the rudder angle around the hover rudder angle. This can be done by introducing the \textit{rudder angle around the hover rudder angle}, denoted by $\uboldthtilde$, as a new variable as follows,
			\begin{align}\label{eq:anglearoundhoverangle}
				\uboldthtilde=\twovector{\deltatildep}{\deltatildes}\equiv\twovector{\deltap-\deltaph}{\deltas-\deltash}\equiv\uboldeta-\uboldth.
			\end{align}
			Fig. \ref{fig:coordinatesystem} visualizes the definition of $\deltatildep$ and $\deltatildes$.
			
\begin{figure*}[htbp]
				\centering
				\includegraphics[width=1.35\columnwidth]{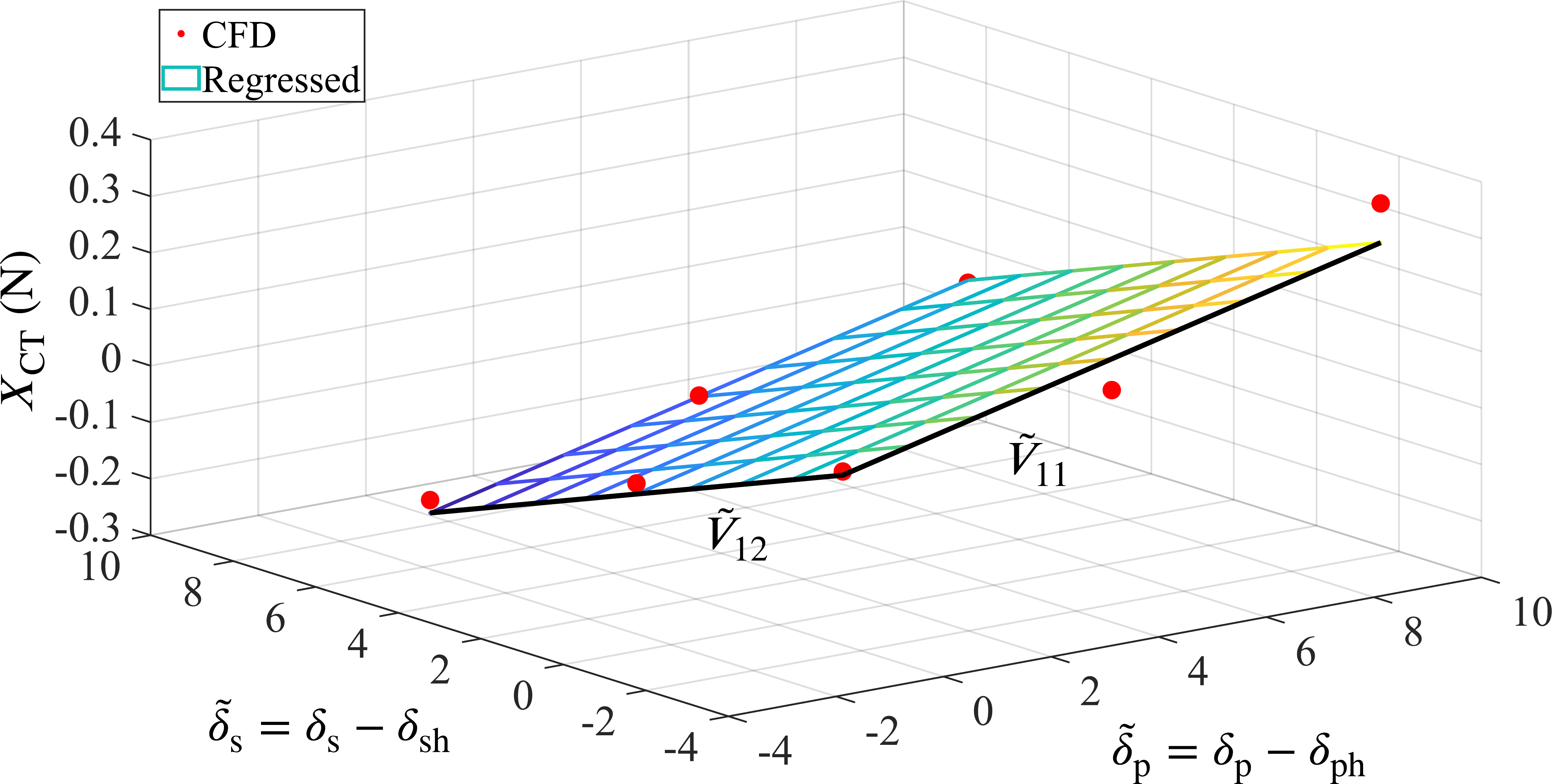}
				\caption{Linearity of $\XCT$ around hover rudder angle.} 
				\label{fig:XCTLinear}
			\end{figure*}

\begin{figure*}[htbp]
				\centering
				\includegraphics[width=1.35\columnwidth]{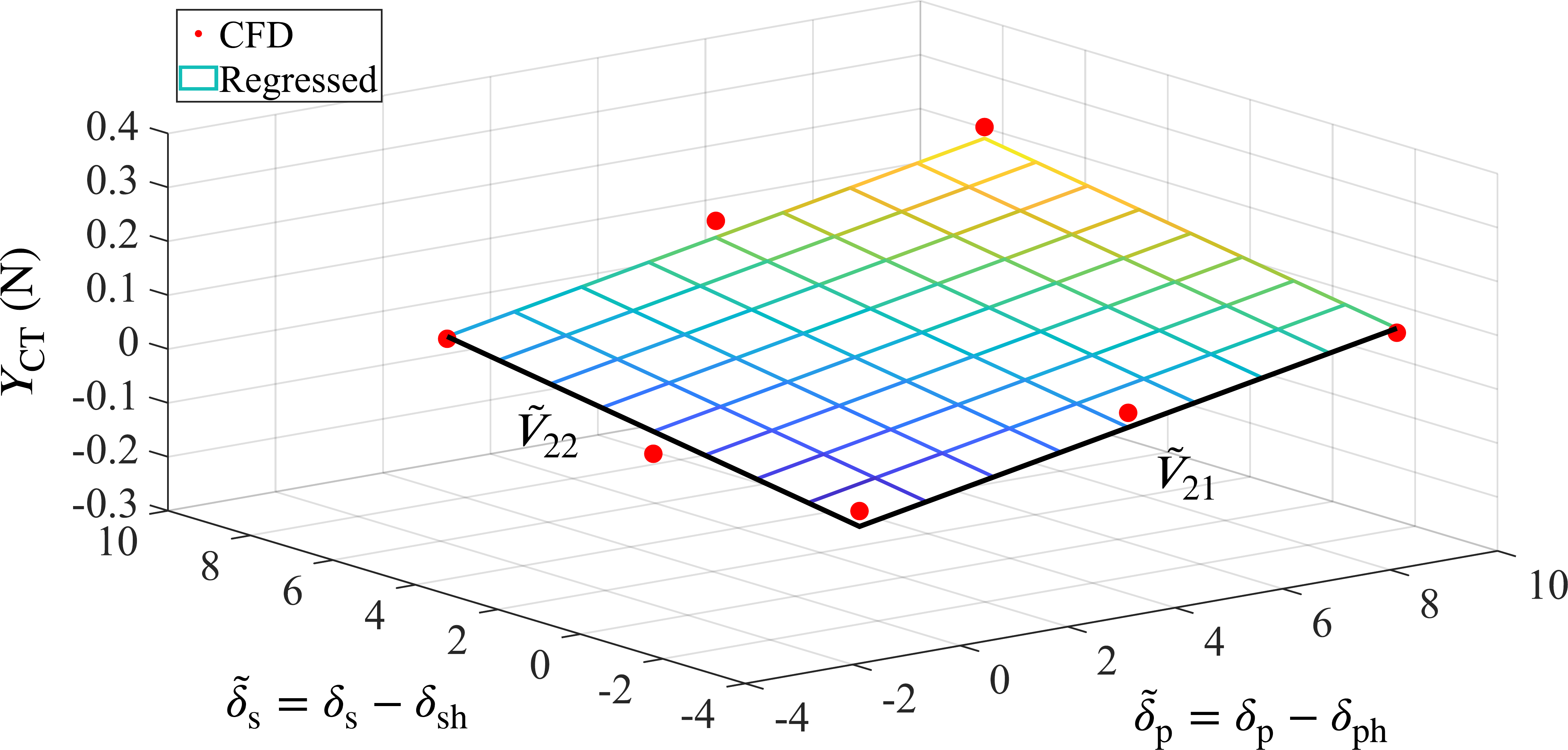}
				\caption{Linearity of $\YCT$ around hover rudder angle.} 
				\label{fig:YCTLinear}
			\end{figure*}

			With (\ref{eq:anglearoundhoverangle}), the $\fbolditcp$ can be excluded such that the relationship in (\ref{eq:regressionsystem}) reduces to
			\begin{align}\label{eq:fequalVtildeu}
				\fboldct=\Vmtxtilde\uboldthtilde.
			\end{align}
			The meaning of (\ref{eq:fequalVtildeu}) is that whenever the total forces $\fboldct=\bm{0}$, the rudders are at the hover rudder angle, i.e., $\uboldeta=\uboldth$. When $\fboldct\neq\bm{0}$, the ship is \textit{slowly} moving with the rudder angle $\uboldeta$ around the hover rudder angle, governed by the linear transformation $\Vmtxtilde$ (\ref{eq:vmtxtilde}). The relationship in (\ref{eq:fequalVtildeu}) is visualized in Fig. \ref{fig:XCTLinear} and Fig. \ref{fig:YCTLinear}.
   
			Further, it is assumed that (\ref{eq:fequalVtildeu}) holds within a wider range of rudder angle: $\deltap\in\lbra{\;-105,\;-60\;}$ and $\deltas\in\lbra{\;60,\;105\;}$. The reason for this is that the VTPS is intended for low-speed operations, thus small angle is immaterial since it may result in a lower drag: higher speed. This assumption is acceptable for low-speed operations as demonstrated in the experiments (see section \ref{Sec:7-Experiment}). The discussion on the applicability of the linear relationship (\ref{eq:vmtxtilde}) for this wider range of rudder angles is presented briefly in appendix \ref{appendixA}.
   
		\subsection{Bow Thruster}
			\label{SubSec:3.2-BowThruster}
			The lateral force from the bow thruster $\YB$ is approximated by
			\begin{align}\label{eq:YB}
				\YB&=\underbrace{\rho D^4 K_{\mathrm{T}}}_{\CB}\underbrace{\nB^2}_{\nBtilde}\nonumber\\
				\YB&=\CB\nBtilde,
			\end{align}
			where $\rho$ is the density of water (kg/m$^3$), $D$ is the diameter of the bow thruster (m), and $K_{\mathrm{T}}$ is the force coefficient that was obtained from a tank test. The new constant: $\CB$ and variable: $\nBtilde$ are introduced  such that $\YB$ is proportional with the square of $\nB$. This "forced linearization" is intentional. One can expect a rather aggressive response from the bow thruster that will be helpful in several situations.  
			
			\subsection{Complete Command-Forces Relationship}\label{SubSec:3.3-ComplCFRel}
			By putting (\ref{eq:fequalVtildeu}) and (\ref{eq:YB}) into one matrix, one can obtain the following command-forces relationship,
			\begin{align}\label{eq:fVuplain}
				\fboldc=\Vmtx\ubold,
			\end{align}
			where the force coefficient matrix $\Vmtx$ is
			\begin{align}\label{eq:vmtxplain}
				\Vmtx = \threethreematrix{\Vtilde_{11}}{\Vtilde_{12}}{0}{\Vtilde_{21}}{\Vtilde_{22}}{0}{0}{0}{\CB}.
			\end{align}
			This $\Vmtx$ maps the modified actuator command $\ubold$,
			\begin{align}\label{eq:actuatorcmdvect2}
				\ubold=\threerowvector{\deltatildep}{\deltatildes}{\nBtilde},
			\end{align}
			to the resulting actuator forces $\fboldc$,
			\begin{align}\label{eq:actuatorforces}
				\fboldc = \threerowvector{\XCT}{\YCT}{\YB},
			\end{align}
			and preserves the origin ($\fboldc=\bm{0}$) at the hover rudder angle $\uboldth$ (\ref{eq:hoverangle}) and $\nB=\;$0.
			
			Notice that the force coefficient matrix $\Vmtx$ is nondiagonal, which is different from the diagonal force coefficient matrix of the usual PID-based DPS \cite{fossen2011handbook,TANNURI2006133,TANNURI20101121}. This is because in the usual DPS, each thruster/rudder can act independently. Each thruster has its own command-thrust relationship, e.g., bow thruster relationship in (\ref{eq:YB}). The absence of one or more thrusters will not affect the other thrusters' forces. However, this is not quite true for the VTPS where the forces are generated through the interaction between the propeller, both rudders, and the hull; all captured when obtaining the $\Vmtx$ matrix.
			
			As a note, there is a possibility for the command-force relationship, i.e., the force coefficient matrix $\Vmtx$ to be extended for a full-scale ship or any arbitrary ship via dimensional analysis. This is discussed briefly in appendix \ref{appendixB}.

\section{Control Forces Allocation} 
    \label{Sec:4-ForcesAlloc}
    A ship requires forces to accelerate. Let $\fboldreq$ denote the required forces that will be distributed/allocated to the available actuators. Also, let $\Xreq$, $\Yreq$, and $\Nreq$ be the required forces to accelerate the surge, sway, and yaw motion, respectively, such that $\fboldreq$ is defined as
    \begin{align}\label{eq:reqforces}
        \fboldreq=\threerowvector{\Xreq}{\Yreq}{\Nreq}.
    \end{align}
    
    The actuator forces $\fboldc$ are allocated to fulfill the required forces $\fboldreq$. From Fig. \ref{fig:coordinatesystem}, the contribution from each actuator to $\fboldreq$ can be expressed as
    \begin{align}\label{eq:reqforcesdistb1}
        \threevector{\Xreq}{\Yreq}{\Nreq}=\threevector{\XCT}{\YCT+\YB}{\YCT\xfR+\YB\xB},
    \end{align}
    where $\xB$ is the longitudinal position of the bow thruster from $\xG$ (see Table \ref{tab:A1-TakaokiPP}). With (\ref{eq:actuatorforces}), the equations in (\ref{eq:reqforcesdistb1}) can be written as the following transformation,
    \begin{align}\label{eq:reqforcesdistb2}
        \fboldreq=\Tmtx\fboldc.
    \end{align}
    where the configuration matrix $\Tmtx$ is as follows,
    \begin{align}\label{eq:configurationmtx}
      \Tmtx=\threethreematrix{1}{0}{0}{0}{1}{1}{0}{\xfR}{x\Bsub}.
    \end{align}
    
    Substitution of the command-forces relationship $\fboldc$ from (\ref{eq:fVuplain}) to (\ref{eq:reqforcesdistb2}) yields a transformation that maps the modified actuator command $\ubold$ (\ref{eq:actuatorcmdvect2}) to the required forces $\fboldreq$ as follows,
    \begin{align}\label{eq:freqTVu}
        \fboldreq=\Tmtx\Vmtx\ubold.
    \end{align}
    Since $\Tmtx$ and $\Vmtx$ are constants, the multiplication is also constant. Let us denote this multiplication as $\Zmtx=\Tmtx\Vmtx$ so that $\Zmtx\invers=\Vmtx\invers\Tmtx\invers$. Then the actuator command that corresponds to the required forces can be computed by
    \begin{align}\label{eq:controlalloc}
        \ubold=\Zmtx\invers\fboldreq.
    \end{align}
    Equation (\ref{eq:controlalloc}) is the main part of the control forces allocation module. Note that the system in (\ref{eq:freqTVu}) can be solved analytically without calculating the inverse numerically. This can be done by first eliminating $\YB$ from (\ref{eq:reqforcesdistb1}) to obtain the expressions for $\YCT$ and $\YB$, then inserting them to (\ref{eq:fVuplain}). Finally, by eliminating either $\deltatildep$ or $\deltatildes$ from the resulting system, one can obtain the expressions for the modified actuator commands.
    
\section{PID Controller}     
    \label{Sec:5-PIDControl}
    Now it is a matter of how to determine the required forces $\fboldreq$. Regardless of the positioning objective (path-following/position-keeping), $\fboldreq$ can be regulated by the traditional decoupled PID controller.

    \begin{figure}[htbp]
        \centering
        \includegraphics[width=0.84\columnwidth]{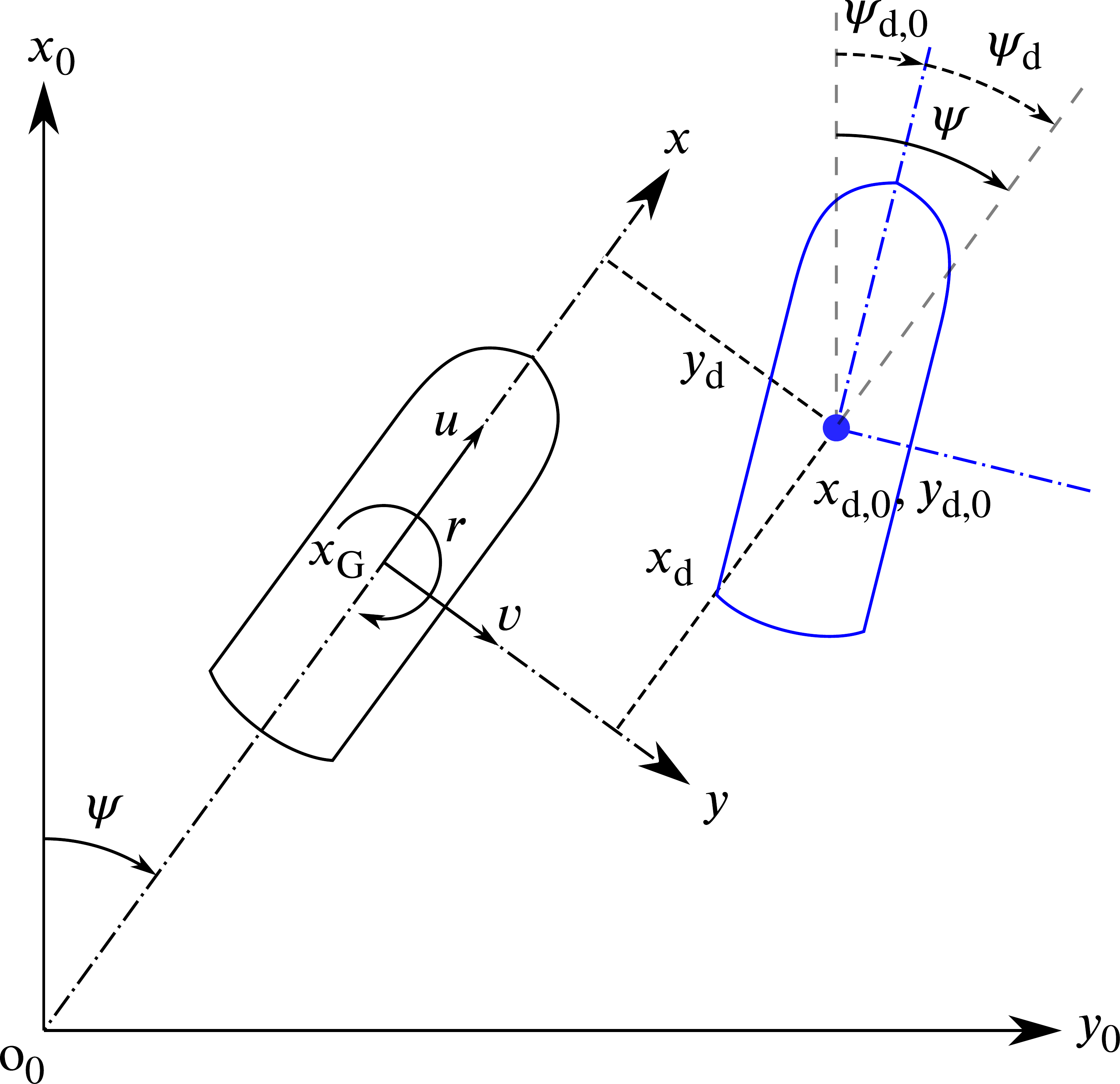}
        \caption{Desired pose: position and yaw angle}. 
        \label{fig:desiredpose}
    \end{figure}
    
    Suppose an earth-fixed coordinate system $\mathrm{o}_0-x_0y_0$ is introduced (see Fig. \ref{fig:desiredpose}). In this coordinate system, the position of $\xG$ is denoted by $x_0$ and $y_0$. Meanwhile, the desired position is denoted by $x\dzerosub$ and $y\dzerosub$. The yaw $\psi$ is defined as the angle between the ship-fixed coordinate system $\xG-xy$ and the earth-fixed coordinate system $\mathrm{o}_0-x_0y_0$. In the same notation, the desired yaw angle is denoted by $\psi\dzerosub$. 
    
    Let us now define the instantaneous pose of the ship in the earth-fixed coordinate system as
    \begin{align}\label{eq:pose}
        \etabold=\threerowvector{x_0}{y_0}{\psi},
    \end{align}
    and the desired pose in the earth-fixed coordinate system as
    \begin{align}\label{eq:desiredpose}
        \etabold\dzerosub=\threerowvector{x\dzerosub}{y\dzerosub}{\psi\dzerosub}.
    \end{align}
    Moreover, let $\vbold$ denote the vector of ship's velocities,
    \begin{align}\label{eq:velovect}
        \vbold=\threerowvector{u}{v}{r},
    \end{align}
    where $u$, $v$, and $r$ denote the surge velocity, sway velocity, and yaw rate, respectively.
    
    Because all the forces (\ref{eq:freqTVu}) are expressed in the ship-fixed coordinate system, the desired pose must also be transformed to the same coordinate system.  From Fig. \ref{fig:desiredpose}, the following transformation can be constructed,
    \begin{align}\label{eq:ezeroRe}
        \ebold_0=\Rpsi\ebold,
    \end{align}
    where $\ebold_0$ is the error of the pose defined in the earth-fixed coordinate system,
    \begin{align}\label{eq:eboldzero}
        \ebold_0 = \threevector{x\dzerosub-x_0}{y\dzerosub-y_0}{\psi\dzerosub-\psi}\equiv\etabold\dzerosub-\etabold,
    \end{align}
    and $\ebold$ is the transformed desired pose,
    \begin{align}\label{eq:transformeddesiredpose}
        \ebold=\threerowvector{x\dsubs}{y\dsubs}{\psi\dsubs},
    \end{align}
    with $x\dsubs$, $y\dsubs$, and $\psi\dsubs$ denote the desired position and yaw, respectively, defined in the ship-fixed coordinate system. As a note, $\ebold$ can also be regarded as the error of the current pose from the desired pose. 
    
    Here the rotation matrix $\Rpsi$ is defined as
    \begin{align}\label{eq:rotmatrix}
        \Rpsi=\threethreematrix{\cospsi}{-\sinpsi}{0}{\sinpsi}{\cospsi}{0}{0}{0}{1}.
    \end{align}
    Because $\Rpsi$ is an orthogonal transformation ($\Rmtx\invers=\Rmtx\transp$), $\ebold$ can be computed from (\ref{eq:ezeroRe}) as
    \begin{align}\label{eq:eRinvezero}
        \ebold=\Rmtx\transp\lpare{\psi}\ebold_0.
    \end{align}
    
    In a simple way, the control objective is to translate $\xG-xy$ over a vector of $x\dzerosub$ and $y\dzerosub$, and rotate it by an angle of $\psi\dzerosub$. All of the transformations are captured in the error $\ebold$ (\ref{eq:eRinvezero}). 
    
    The error $\ebold$ should be stabilized at zero by controlling the required forces $\fboldreq$ with the following decoupled PID controller,
    \begin{subequations}\label{eq:PIDcontrollerexpand}
        \begin{align}
            \forcePID{X}{x}{u}\\
            \forcePID{Y}{y}{v}\\
            \forcePID{N}{\psi}{r}.
        \end{align}
    \end{subequations}
    In a compact form, (\ref{eq:PIDcontrollerexpand}) can be expressed as
    \begin{align}\label{eq:PIDcompact}
        \fboldreq = \Kmtx{P}\ebold+\Kmtx{I}\int\ebold\,\dt-\Kmtx{D}\vbold,
    \end{align}
    where 
    \begin{subequations}
        \begin{align}\label{eq:PIDgains}
			\Kmtx{P}&=\mathrm{diag}\lpare{\;K\PIDcoeff{PX},\;K\PIDcoeff{PY},\;K\PIDcoeff{PN}\;}\\
			\Kmtx{I}&=\mathrm{diag}\lpare{\;K\PIDcoeff{IX},\;K\PIDcoeff{IY},\;K\PIDcoeff{IN}\;}\\
			\Kmtx{D}&=\mathrm{diag}\lpare{\;K\PIDcoeff{DX},\;K\PIDcoeff{DY},\;K\PIDcoeff{DN}\;}.
        \end{align}
    \end{subequations}
    Here, $\Kmtx{P}, \Kmtx{I}, \Kmtx{D}\in \mathbb{R}^{3\times3}$ are positive diagonal gain matrices for the proportional, integral, and derivative term, respectively. The use of velocities $\vbold$ instead of the time derivative of the error ($\mathrm{d}\ebold/\mathrm{d}t$) is to reduce derivative kicks that could happen in a path-following operation (due to setpoint switch). 
    
    Equation (\ref{eq:PIDcompact}) is the essence of the control allocation module. It will determine the required forces $\fboldreq$ and send them to the control forces allocation module (\ref{eq:controlalloc}) to obtain the corresponding modified actuator command $\ubold$. 
    
\section{Desired Pose Selection for the VTPS}
	\label{Sec:6-PathRef}
	In the usual automatic maneuvering operation, a path or a trajectory serves as the reference or guide for the operation. A path is a geometric entity made up of a set of waypoints. Yaw angle can be assigned at each waypoint such that the path is smoothed. When the path is scheduled, i.e., constrained by time, it is termed a trajectory. The readers are referred to \cite{ZHOU2020107043}, \cite{Vagale2021}, and \cite{MPCsurvey} for a survey on how to plan such a path/trajectory.

	In this article, several reference trajectories were generated via an OCP-based (optimal control problem) \textit{offline} trajectory planner presented in \cite{maki2020pt1,Maki2020application}, and \cite{Miyauchi2021planner}. One reference trajectory consists of time-scheduled positions, yaw angle, and velocities. Because the VTPS is intended for a path-following operation, it concerns only the pose: positions and yaw angle. These reference poses will be the setpoints in the controller module (\ref{eq:PIDcompact}). Be advised that the term \textit{reference path} also refers to the reference pose. Also note that the terms \textit{reference} and \textit{desired} are equivalent and will be used interchangeably.

	The following passages explain the basis of the reference module: how to select a desired pose. Suppose $\Pmtx\in\mathbb{R}^{q\times 3}$ is an array which rows are the time-ordered reference pose,
	\begin{align}\label{eq:Pmtxpose}
		\Pmtx=\fourrowvector{\etatilde\dzerosub^{1}}{\etatilde\dzerosub^{2}}{\dots}{\etatilde\dzerosub^q},
	\end{align}
	where $q$ is the number of the discretizations of the reference trajectory. Each discretization is called a waypoint. Array $\Pmtx$ makes up a database of $q$ waypoints with the corresponding reference pose assigned at each waypoint.

	Let $k$ be the index of the waypoint such that $k=1,2,\dots,q$. The index of the nearest waypoint, denoted by $j$, from the instantaneous pose of the ship can be determined by
	\begin{align}\label{eq:nearestindex}
		j = \argmin_k \; L\lpare{k,\etabold}.
	\end{align}
	The function $L$ is a weighted norm as follows,
	\begin{align}
		L\lpare{k,\etabold} &= \norm{\etabold-\etatilde\dzerosub^k}_\Wmtx \nonumber \\
		&=\lpare{\etabold-\etatilde\dzerosub^k}\transp\Wmtx\lpare{\etabold-\etatilde\dzerosub^k}\label{eq:Lfunction},
	\end{align}
	where $\Wmtx\in\mathbb{R}^{3\times 3}$ is a positive diagonal weight matrix,
	\begin{align}\label{eq:diagweight}
		\Wmtx=\mathrm{diag}\lpare{\;1,\;1,\;0\;},
	\end{align}
	such that only the positions are considered.

	The next waypoint, i.e., the desired pose can be selected from the database $\Pmtx$ as
	\begin{align}\label{eq:nextwaypoint}
		\etabold\dzerosub\longleftarrow\etatilde\dzerosub^{i},
	\end{align}
	where the index of the desired waypoint $i$ is $\Delta k$ steps ahead from the nearest waypoint $j$, or
	\begin{align}
		i = j + \Delta k.
	\end{align}
	In other words, take the $i$-th row of $\Pmtx$ as the setpoints for the controller (\ref{eq:PIDcompact}). Note that $\Delta k$ should be chosen accordingly. As a rule of thumb, large $\Delta k$ is good for a straight path while small $\Delta k$ is good for a curved path.

\section{Combined Automatic Docking and Position-keeping Scale Model Experiments}
    \label{Sec:7-Experiment}
    As already mentioned in the introduction, the purpose of the study is to test the VTPS directly in scale model experiments. These experiments should be done when the wind is calm with no/weak gusts. There were a few good opportunities to conduct the experiments within the first half of the year, which in Japan is known for its moderate-strong wind.
    
    In one single run, two experiments were conducted:
    \begin{enumerate}
        \item Automatic docking that can be divided into two stages: transition and docking. The transition stage is when the ship moves from a point near the open sea to a receiving point. Maneuver from the receiving point and stop at a final docking point is the docking stage.
        \item Position-keeping: stabilize the ship's pose at the final docking point. \end{enumerate}
    
        \subsection{Experiment Setup}
			\label{SubSec:7.1-ExpSetup}
            The scale model ship is equipped with a computer that runs on Ubuntu with robotic operating system (ROS). The ship acts as a master node in the ROS environment. The main modules of the VTPS are installed in a land computer as a \MATLAB Simulink node. 
            
            The ship publishes the measurements from the onboard sensors (see Table \ref{tab:OverviewComputer}) to the VTPS in the land computer via a wireless connection. Based on the received state measurements, the VTPS will synthesize the actuator commands and send them to the ship. 
            
            Together, the ship and the \MATLAB Simulink in the land computer create a ROS network that provides communication between the sensors, actuators, and controller. The update frequency of this communication is 10 Hz; updated every 0.1 s.
            
            \begin{table}[h]
                \caption{State measurements.}\label{tab:OverviewComputer}
				\begin{tabular*}{\columnwidth}{ @{\extracolsep{\fill}} l l }
					\hline\noalign{\smallskip}
					States & Onboard sensors  \\ 
					\noalign{\smallskip}\hline\noalign{\smallskip}
					$x_0$ (m) & \multirow{3}{*}{3 GNSS units, Magellan Systems Japan, Inc.} \\
					$y_0$ (m) &  \\
					$u$ (m/s) & \multirow{1}{*}{MJ-3021-GM4-QZS-EVK and MJ-3008-GM4-QZS} \\
					$v$ (m/s) &  \\
					\noalign{\smallskip}\hline\noalign{\smallskip}
					$\psi$ (rad) & Fiber optic gyro, Japan Aviation Electronics \\
					$r$ (rad/s) & JG-35FD \\ 
					\noalign{\smallskip}\hline
				\end{tabular*}
            \end{table} 
        
        \subsection{Automatic Docking Experiments}
			\label{SubSec:7.2-AutoDockExp}
			At first, the ship is manually controlled to a desired initial point. The experiment starts at this initial point. At the same time, an array $\Pmtx$ that contains the reference poses (waypoints) is selected based on this initial point.
			
			During the first stage: the transition stage, the VTPS is inactive. Instead, a simple PID controller with line-of-sight (LOS) guidance \cite{FOSSEN2003211} is active to control the rudders such that the ship follows the desired path. The propeller is constant in forward mode and the rudders are allowed to move at their full range (see Table \ref{tab:actuatorlimittransition}).
			
			\begin{table}[h]
				\caption{Actuators range in the transition stage.}\label{tab:actuatorlimittransition}
				\begin{tabular*}{\columnwidth}{ @{\extracolsep{\fill}} l l }
					\hline\noalign{\smallskip}
					Actuators & Range \\ 
					\noalign{\smallskip}\hline\noalign{\smallskip}
					Port rudder (deg) & $\deltap\in\lbra{\;-105,\;35\;}$ \\
					Starboard rudder (deg) & $\deltas\in\lbra{\;-35,\; 105 \;}$ \\
					Bow thruster (rps) & $\nB=0$ (Off) \\
					Propeller (rps) & $n=10$ or $n=7$ \\
					\noalign{\smallskip}\hline
				\end{tabular*}
			\end{table}
			
			The ship enters the next stage: the docking stage when it is in the proximity of a predefined switching (receiving) point. In this docking stage, the VTPS is active. The VTPS slows the ship down and controls the ship to follow the remaining path before finally stop at the final docking point. 
			
			The propeller is constant in forward mode, while the ranges of the rudders are limited based on the assumption in sub-subsection \ref{SubSubSection:3.1.2-LinearHVR}. The ranges of the actuators in this stage are summarized in Table \ref{tab:actuatorlimitdocking}. Moreover, saturation values for the required forces and the PID gains in the VTPS are given in Table \ref{tab:freqlimit} and Table \ref{tab:PIDgains}, respectively. 

            The PID gains are determined from extensive trials in the experiment pond. The chosen gains are those that balance the performance during both docking and positioning-stage. Proportional gains led to unnecessary oscillations in the position-keeping stage but were helpful in the docking stage. The derivative gains, on the other hand, were very crucial in the position-keeping stage but reduced the performance in the docking stage. The integral is set as low as possible considering that the gusts in the pond usually blow within a very short time. 

			\begin{table}[h]
				\caption{Actuators range for the VTPS (docking stage and position-keeping).}\label{tab:actuatorlimitdocking}
				\begin{tabular*}{\columnwidth}{ @{\extracolsep{\fill}} l l }
					\hline\noalign{\smallskip}
					Actuators & Range \\ 
					\noalign{\smallskip}\hline\noalign{\smallskip}
					Port rudder (deg) & $\deltap\in\lbra{\;-105,\;-60\;}$ \\
					Starboard rudder (deg) & $\deltas\in\lbra{\;60,\; 105 \;}$ \\
                    Rudder rate (deg/s) & $\approx 23$\\
					Bow thruster (rps) & $\nB\in\lbra{\;-27,\; 27 \;}$ \\
					Propeller (rps) & $n=10$ \\
					\noalign{\smallskip}\hline
				\end{tabular*}
			\end{table}    
			
			\begin{table}[h]
				\caption{Saturation values for the required forces $\fboldreq$ from the PID controller in the VTPS (docking stage and position-keeping).}\label{tab:freqlimit}
				\begin{tabular*}{\columnwidth}{ @{\extracolsep{\fill}} l l }
					\hline\noalign{\smallskip}
					Required forces & Range \\ 
					\noalign{\smallskip}\hline\noalign{\smallskip}
					Surge (N) & 
					$-1.5\leq\Xreq \leq 0.8$\\
					Sway (N) & $-1.0\leq\Yreq\leq 1.0$\\
					Yaw moment (N-m) & $-1.7\leq \Nreq\leq 1.5$ \\
					\noalign{\smallskip}\hline
				\end{tabular*}
			\end{table}       
			
			\begin{table}[h]
				\caption{PID gains in the VTPS (docking stage and position-keeping).}\label{tab:PIDgains}
				\begin{tabular*}{\columnwidth}{ @{\extracolsep{\fill}} l l }
					\hline\noalign{\smallskip}
					Terms & Values \\ 
					\noalign{\smallskip}\hline\noalign{\smallskip}
					Proportional & $\Kmtx{P}=\mathrm{diag}\lpare{\;4,\;4,\;4\;}$\\
					Integral & $\Kmtx{I}=\mathrm{diag}\lpare{\;0.01,\;0.01,\;0.001\;}$\\
					Derivative & $\Kmtx{D}=\mathrm{diag}\lpare{\;25,\;25,\;30\;}$\\
					\noalign{\smallskip}\hline
				\end{tabular*}
			\end{table} 
			
		\subsection{Position-keeping Experiments}
			\label{SubSec:7.3-PosKeepExp}
			After the ship arrives and stops at the final docking point, the VTPS has to stabilize the pose of the ship, i.e., the desired pose remains unchanged or $\etabold\dzerosub=\etatilde\dzerosub^{q}$ (see section \ref{Sec:6-PathRef}). The ship must stay stationary, and when the gust is strong, the VTPS should return the ship to the desired pose. This position-keeping experiment lasts for an extended period. Note that the VTPS still uses the same settings as in the docking stage (see Table \ref{tab:actuatorlimitdocking} to Table \ref{tab:PIDgains}).

		\subsection{Results and Discussion}
			\label{SubSec:7.4-ResultsExp}
			This article presents three experiments that were conducted from March to April 2022, each is referred to as experiment A, B, and C, respectively. In experiment A, the propeller was 10 rps all the time. In experiment B, for a reason that will be explained later, the propeller was set to 7 rps in the transition stage and 10 rps for the rest of the experiment. Lastly, experiment C was conducted to test the VTPS for a different type of path: 180 degree/U-turn docking, where the propeller was 10 rps all the time. Note that the propeller was in forward mode only.

			The discussion that follows will focus mainly on the performance of the VTPS in the docking stage and position-keeping, hence the limited information on the first stage that is shown in the figures. Moreover, the readers can refer to the animations of each experiment in the electronic supplementary materials (digital/web version only).

			\subsubsection{Experiment A}
				\label{SubSubSection:7.4.1-FirstExp}
\begin{figure}[htbp]
					\centering
					\includegraphics[width=\columnwidth]{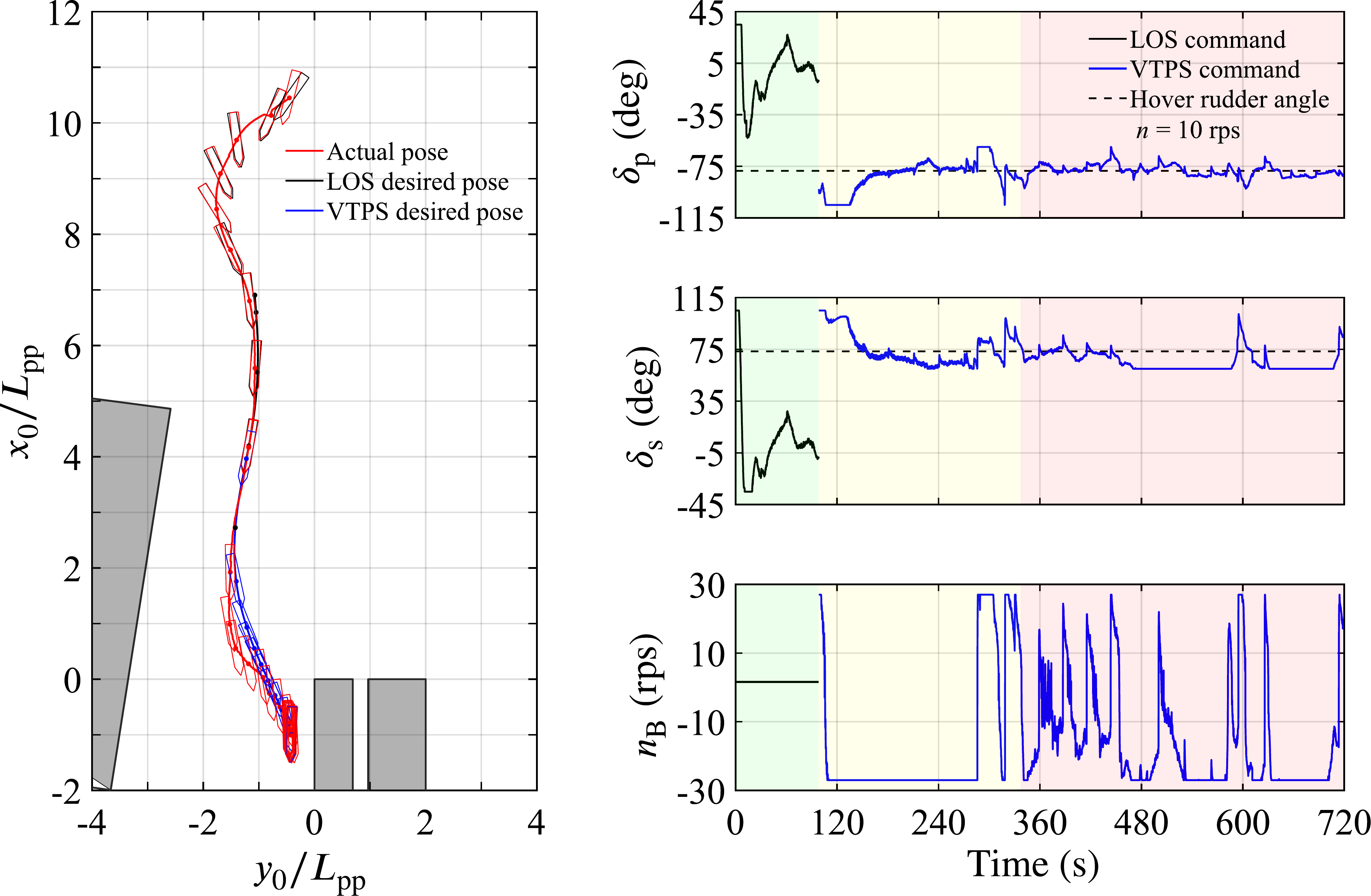}
					\caption{Visualization and time histories of control commands in experiment A.} 
					\label{fig:visual1exp}
				\end{figure}

\begin{figure}[htbp]
					\centering
					\includegraphics[width=\columnwidth]{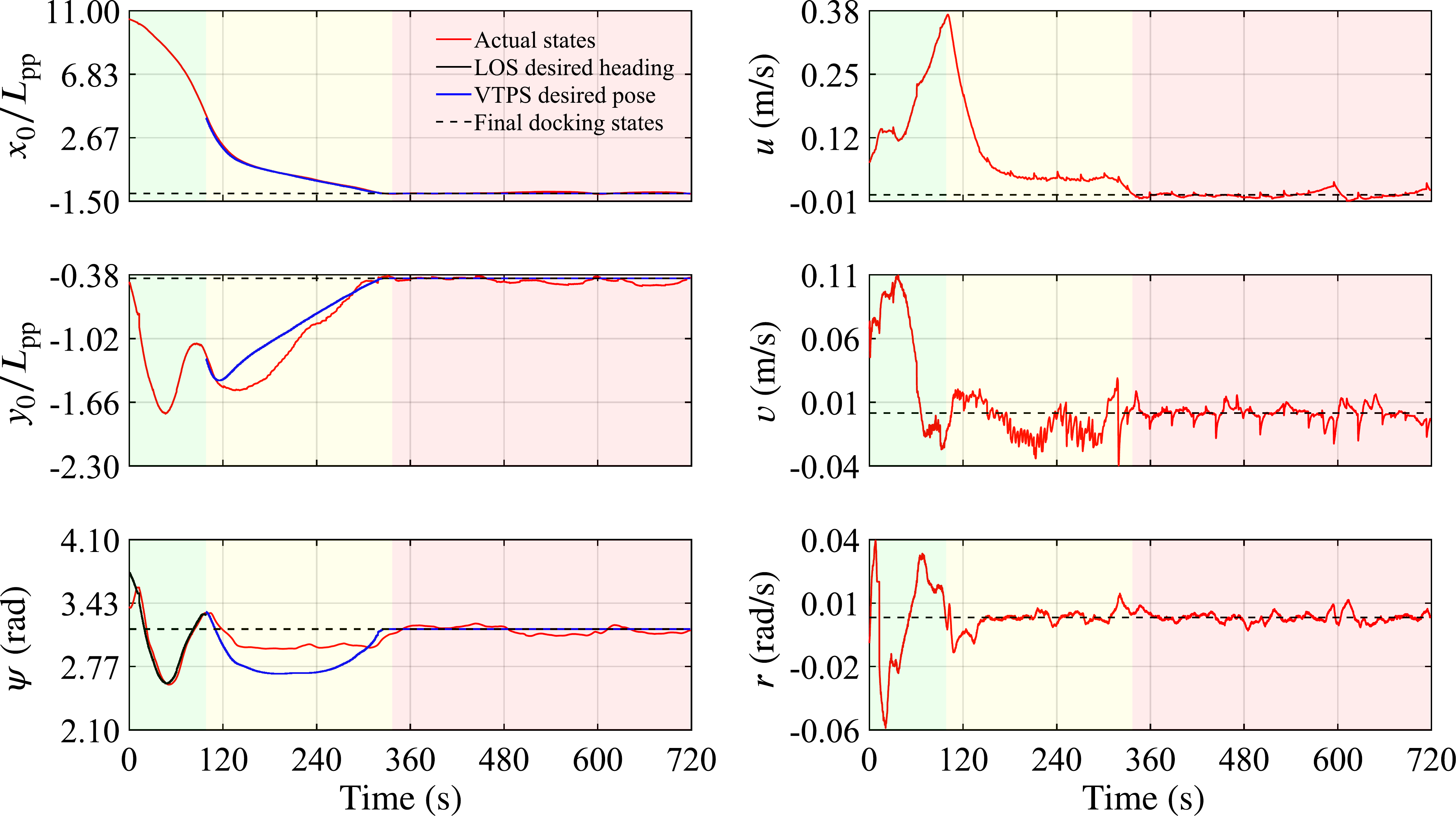}
					\caption{Time histories of ship's states in experiment A.} 
					\label{fig:states1exp}
				\end{figure}

				Experiment A was conducted in a calm wind with 0.3 m/s to 2.0 m/s gusts. The visualization and the time history of the control commands are shown in Fig. \ref{fig:visual1exp}. Meanwhile, the time histories of the states are shown in Fig. \ref{fig:states1exp}. To distinguish each stage: the transition stage, the docking stage, and the position-keeping stage are shaded with green, yellow, and red, respectively.

				\begin{figure}[htbp]
					\centering
					\includegraphics[width=\columnwidth]{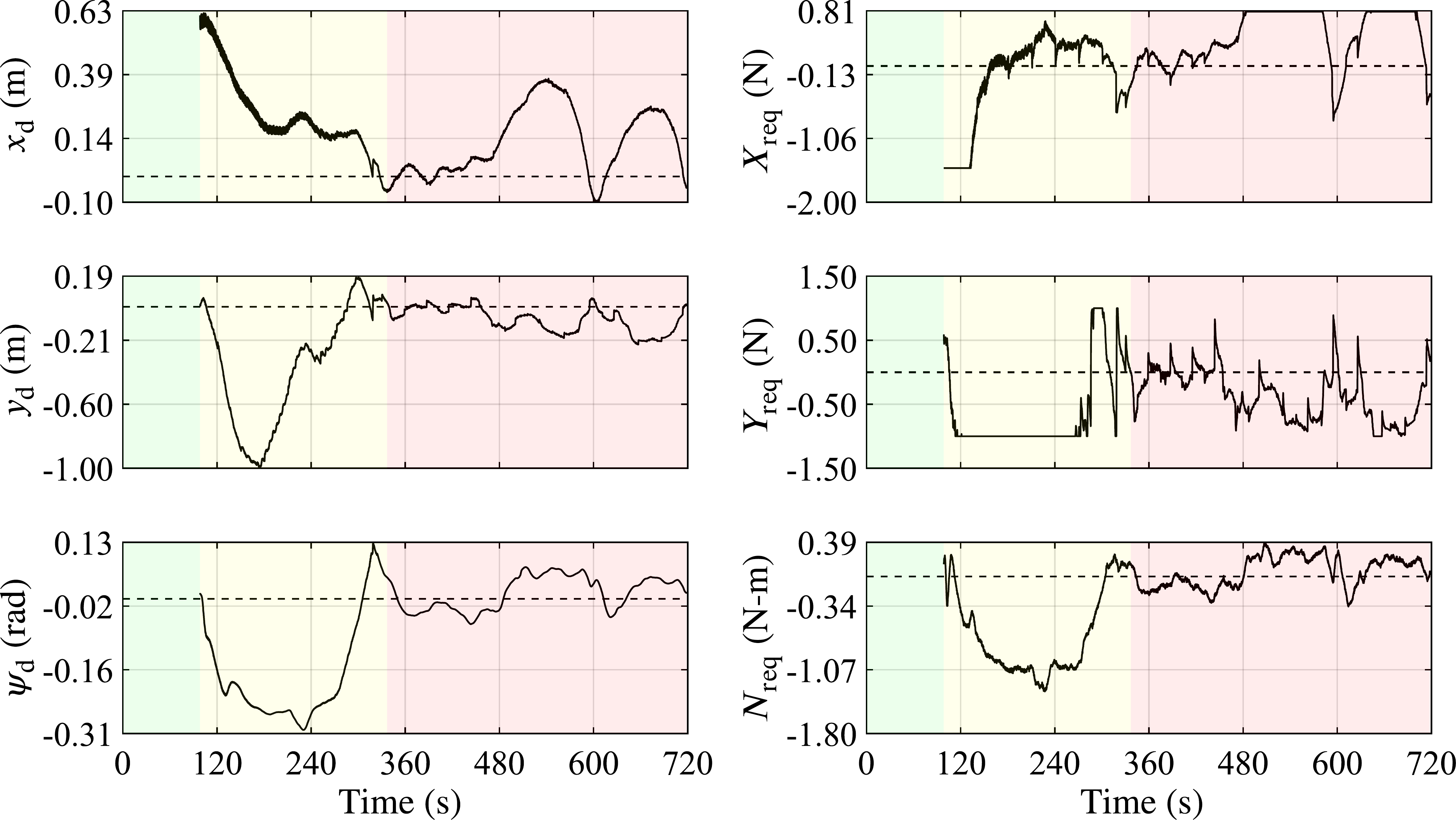}
					\caption{Desired pose in ship-fixed coordinate system and required forces from the VTPS (experiment A).} 
					\label{fig:desiredposeandforces1exp}
				\end{figure}

				One can observe that in the first stage (shaded with green), when the ship was at the desired path, the surge velocity $u$ was increasing because the rudders were zero (no error). At a constant $n=\;$10 rps, the acceleration was quite fast as can be seen in Fig. \ref{fig:states1exp}. The ship accelerated from $u\approx\;$0.10 m/s to $u\approx\;$0.37 m/s in less than two minutes. 

				When the ship arrived at the switching point, the VTPS was activated, and thus the docking stage began (shaded with yellow). The VTPS performed very well to decelerate the ship from $u\approx\;$0.37 m/s to $u\approx\;$0.03 m/s in just 40 s; equivalent to a crash-stopping. This shows the excellent ability of the VecTwin rudder in doing a crash-stopping at a constant forward propeller revolution as investigated in \cite{1997197HAMAMOTO}.

				When crash-stopping (or sudden deceleration in general), the ship deviated from the path. However, the VTPS was able to return the ship back to the desired path. The VTPS synthesized the corresponding actuator commands such that the ship followed the remaining desired poses and finally stopped at the final docking point, i.e., $u=v\approx\;$0 m/s and $r\approx\;$0 rad/s. 

				The remaining time is the position-keeping experiment, shaded with red. The VTPS gave satisfactory performances in stabilizing the error of the ship's pose at the final docking point for about six minutes straight. This can be seen from Fig. \ref{fig:states1exp} where the ship's states (plotted in solid red lines) are stabilized around the desired values (plotted in dashed black lines). This can also be observed from Fig. \ref{fig:desiredposeandforces1exp} that shows the error $\ebold=\threerowvector{x\dsubs}{y\dsubs}{\psi\dsubs}$ and the corresponding required forces $\fboldreq=\threerowvector{\Xreq}{\Yreq}{\Nreq}$. 

\begin{figure}[htbp]
					\centering
					\includegraphics[width=0.7\columnwidth]{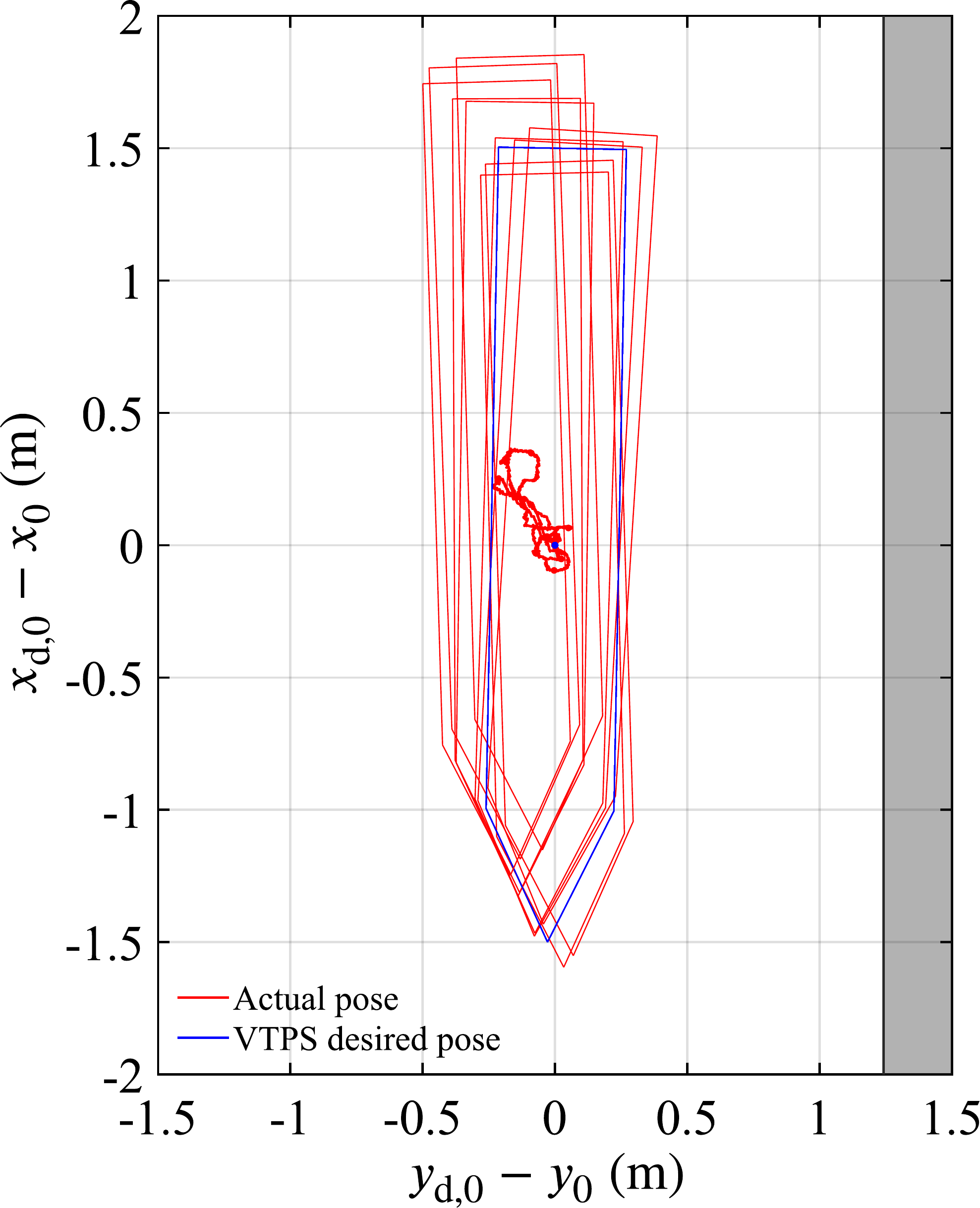}
					\caption{Pose of the ship relative to the desired pose during position-keeping in experiment A.} 
					\label{fig:poskeep1exp}
				\end{figure}

				More clearly, the deviation/error from the desired pose in the earth-fixed coordinate system $\lpare{\ebold_0}$ during the position-keeping experiment is visualized in Fig. \ref{fig:poskeep1exp}. The gusts during the experiment tend to push the ship in the stern-starboard direction. Nevertheless, the VTPS was still able to stabilize the ship back to the desired pose.
				
			\subsubsection{Experiment B}
			    \label{SubSubSection:7.4.2-SecondExp}
    			Experiment B was conducted in a calm wind with 0.1 m/s to 2.0 m/s gusts. In this experiment, the propeller revolution in the first stage was reduced to 7 rps instead of 10 rps. In the remaining stages, the propeller revolution was 10 rps (see Table \ref{tab:actuatorlimitdocking}). Moreover, the switching point was moved farther from the final docking point. This resulted in the transition stage that was much shorter than that in experiment A. Doing this also gave more chances to test the performance of the VTPS in the docking stage.  
    			 
    			The reason for the above treatments is to reduce the excessive acceleration of the ship that is likely to happen in the transition stage. Such treatments are also expected to reduce the deviation from the desired path that was observed during experiment A when decelerating (switching to the docking stage). 
    			    			
    			\begin{figure}[htbp]
					\centering
					\includegraphics[width=\columnwidth]{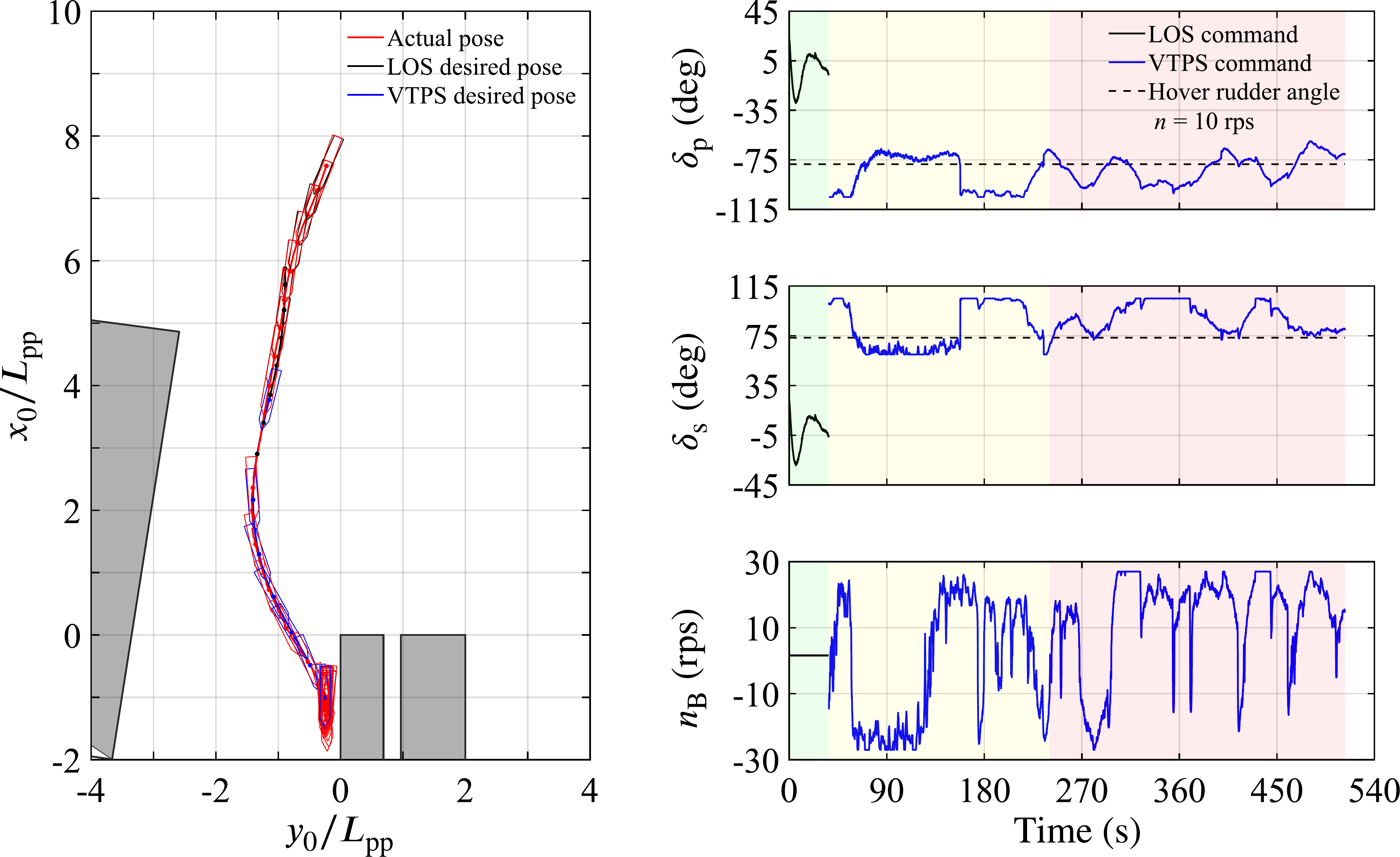}
					\caption{Visualization and time histories of control commands in experiment B, with $n=\;$7 rps in the transition stage and $n=\;$10 rps in other stages.} 
					\label{fig:visual2exp}
				\end{figure}
				
				\begin{figure}[htbp]
					\centering
					\includegraphics[width=\columnwidth]{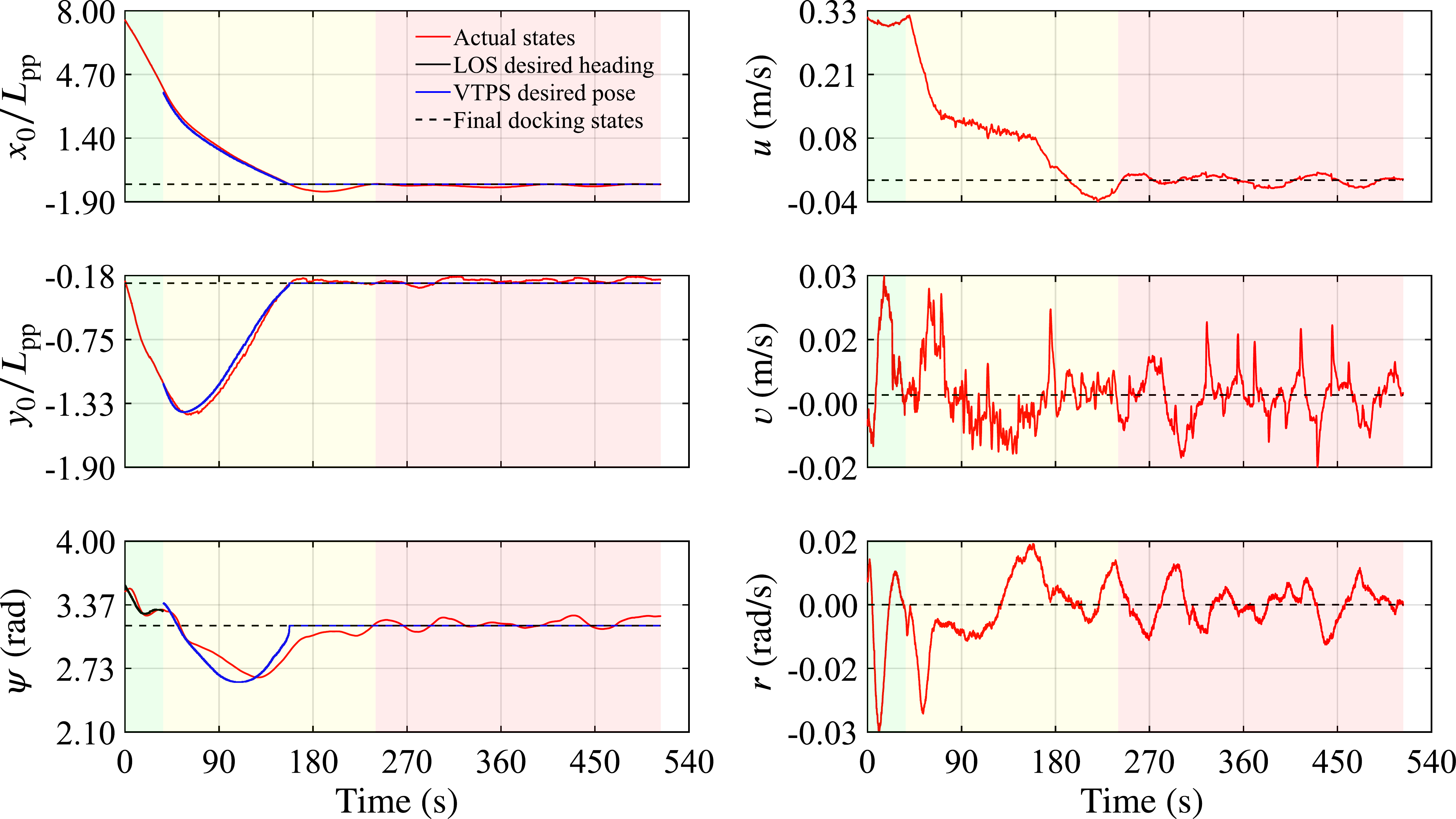}
					\caption{Time histories of ship's states in experiment B.} 
					\label{fig:states2exp}
				\end{figure}

                \begin{figure}[htbp]
					\centering
					\includegraphics[width=\columnwidth]{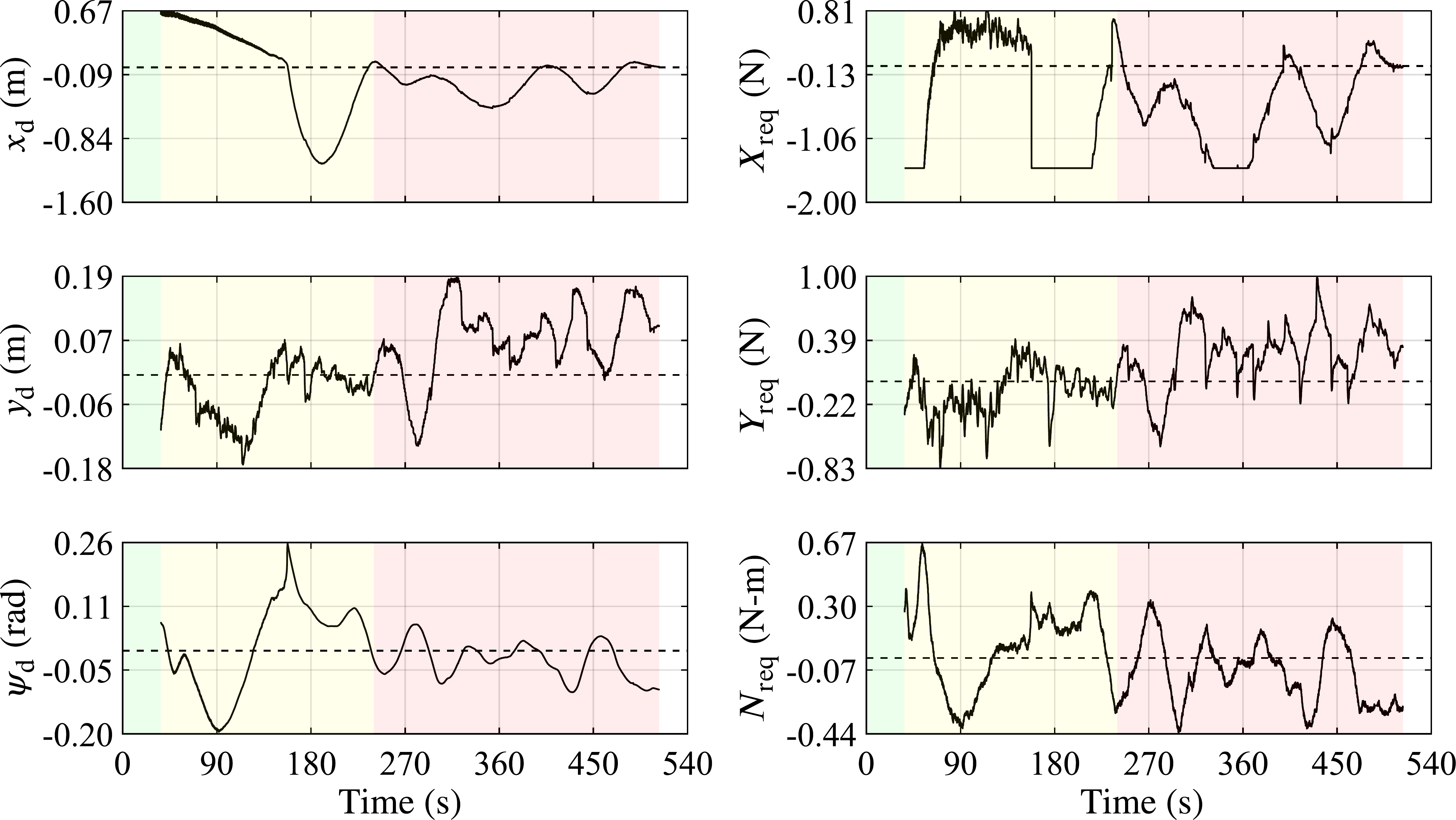}
					\caption{Desired pose in ship-fixed coordinate system and required forces from the VTPS (experiment B).} 
					\label{fig:desiredposeandforces2exp}
				\end{figure}
				
    			As expected, the acceleration in the transition stage was minimized. The ship's surge velocity $u$ was maintained at around 0.30 m/s as shown in Fig. \ref{fig:states2exp} (shaded with green). The surge velocity was also reduced gradually in the docking stage (shaded with yellow). More importantly, the tendency to rotate during deceleration was minimized. Thus, the deviation from the desired path was also minimized as can be observed in Fig. \ref{fig:visual2exp}. 
    			
    			However, when approaching the end of the docking stage, the ship overshot for about 1.15 m; approximately $2/5\Lpp$. Nevertheless, the VTPS controlled the ship to slowly reverse and finally stop at the final docking point. This overshoot is directly related to the reference (setpoint) module (see section \ref{Sec:6-PathRef}). This issue will be improved in future occasions.
    			
\begin{figure}[htbp]
					\centering
					\includegraphics[width=0.7\columnwidth]{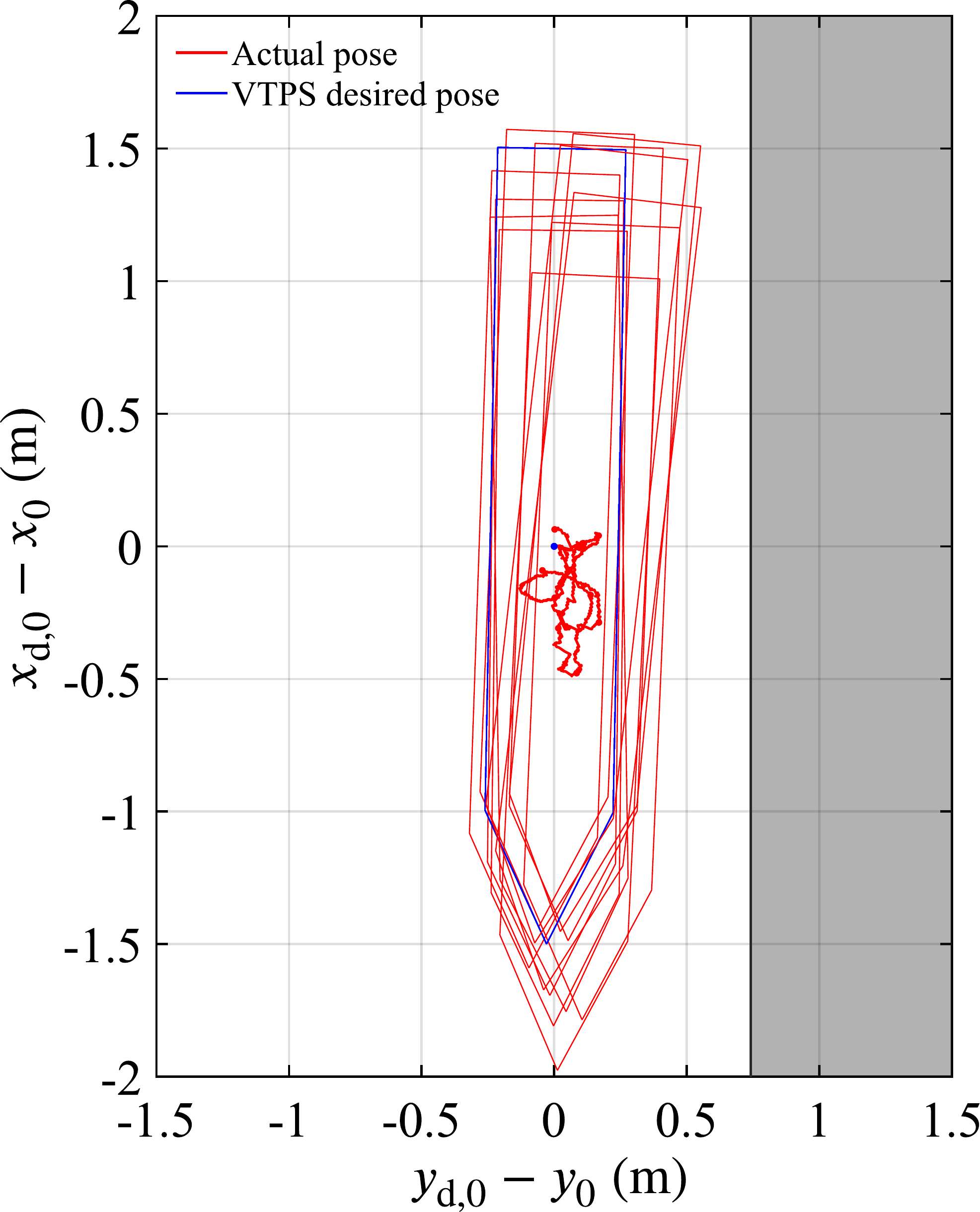}
					\caption{Pose of the ship relative to the desired pose during position-keeping in experiment B.} 
					\label{fig:poskeep2exp}
				\end{figure}

    			For position-keeping, the VTPS performed consistently in stabilizing the pose of the ship at the final docking point for about five minutes. This is shown in Fig. \ref{fig:desiredposeandforces2exp} and visualized in Fig. \ref{fig:poskeep2exp}. The gusts tend to push the ship in the bow-port direction which caused a maximum deviation of approximately 0.5 m; around 0.17$\Lpp$.
				
			\subsubsection{Experiment C}
			    \label{SubSubSection:7.4.3-ThirdExp}
\begin{figure}[htbp]
					\centering
					\includegraphics[width=\columnwidth]{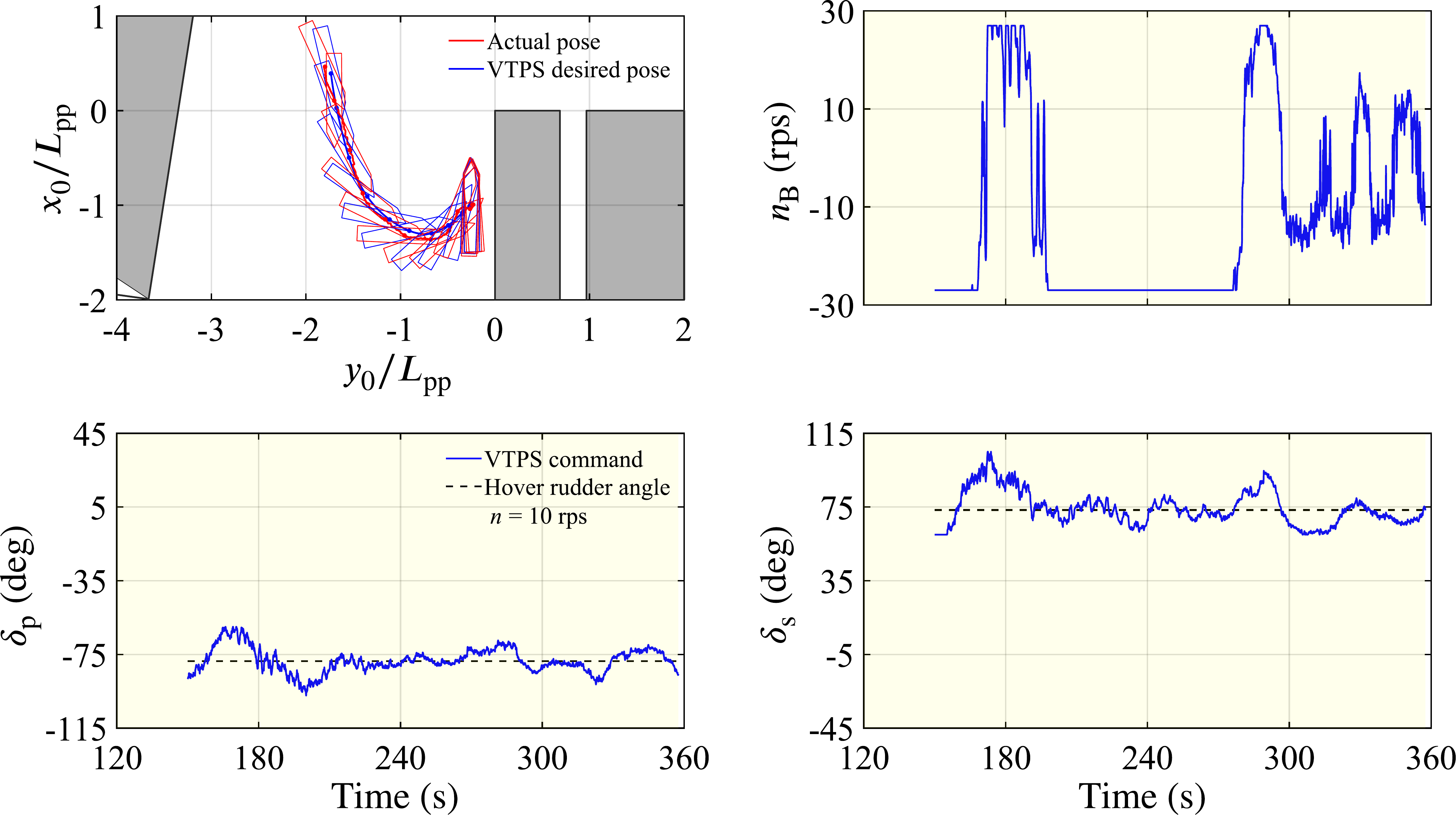}
					\caption{Visualization and time histories of control commands in experiment C.} 
					\label{fig:visual3exp}
				\end{figure}

\begin{figure}[htbp]
					\centering
					\includegraphics[width=\columnwidth]{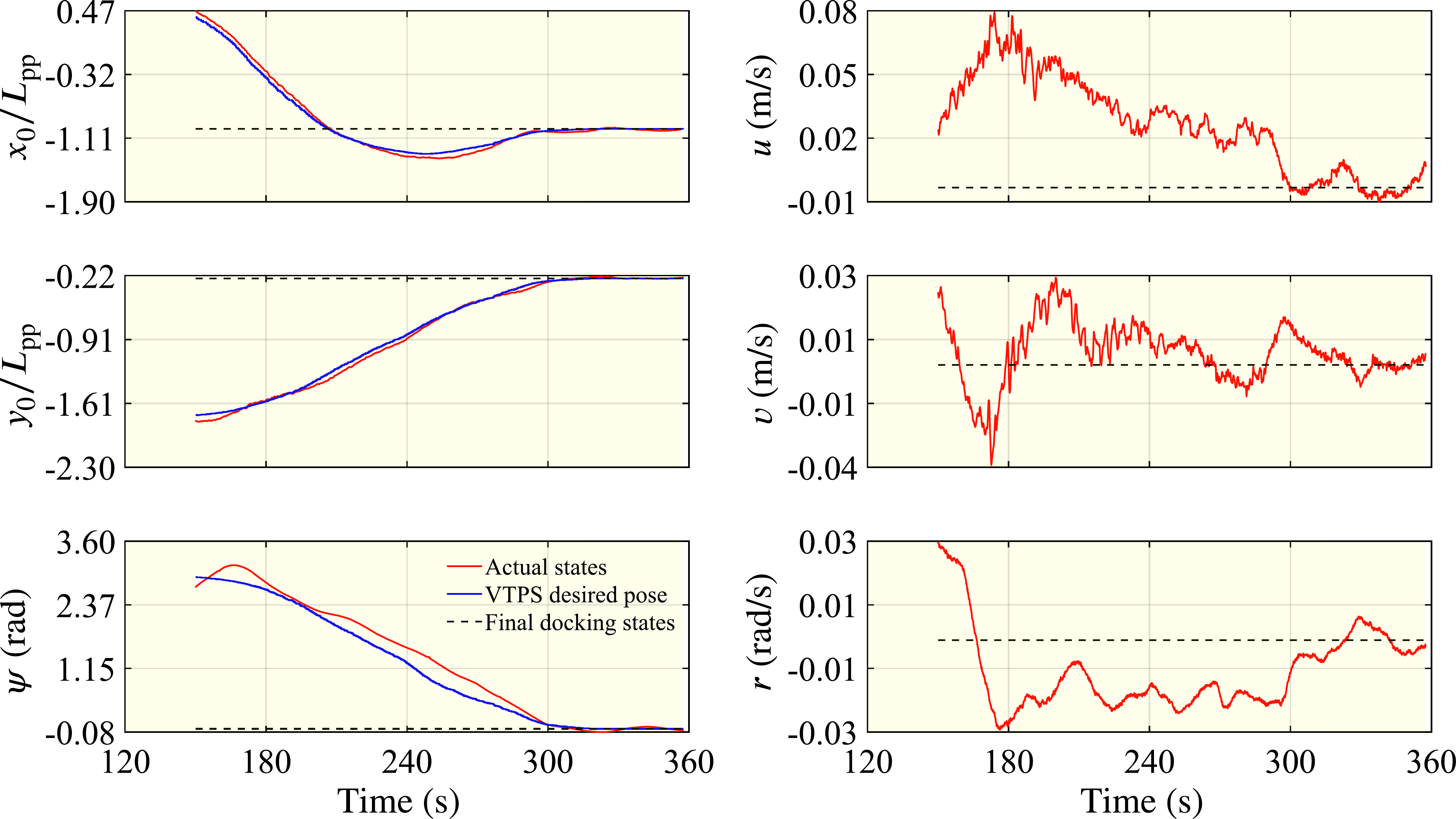}
					\caption{Time histories of ship's states in experiment C.} 
					\label{fig:states3exp}
				\end{figure}

\begin{figure}[htbp]
					\centering
					\includegraphics[width=\columnwidth]{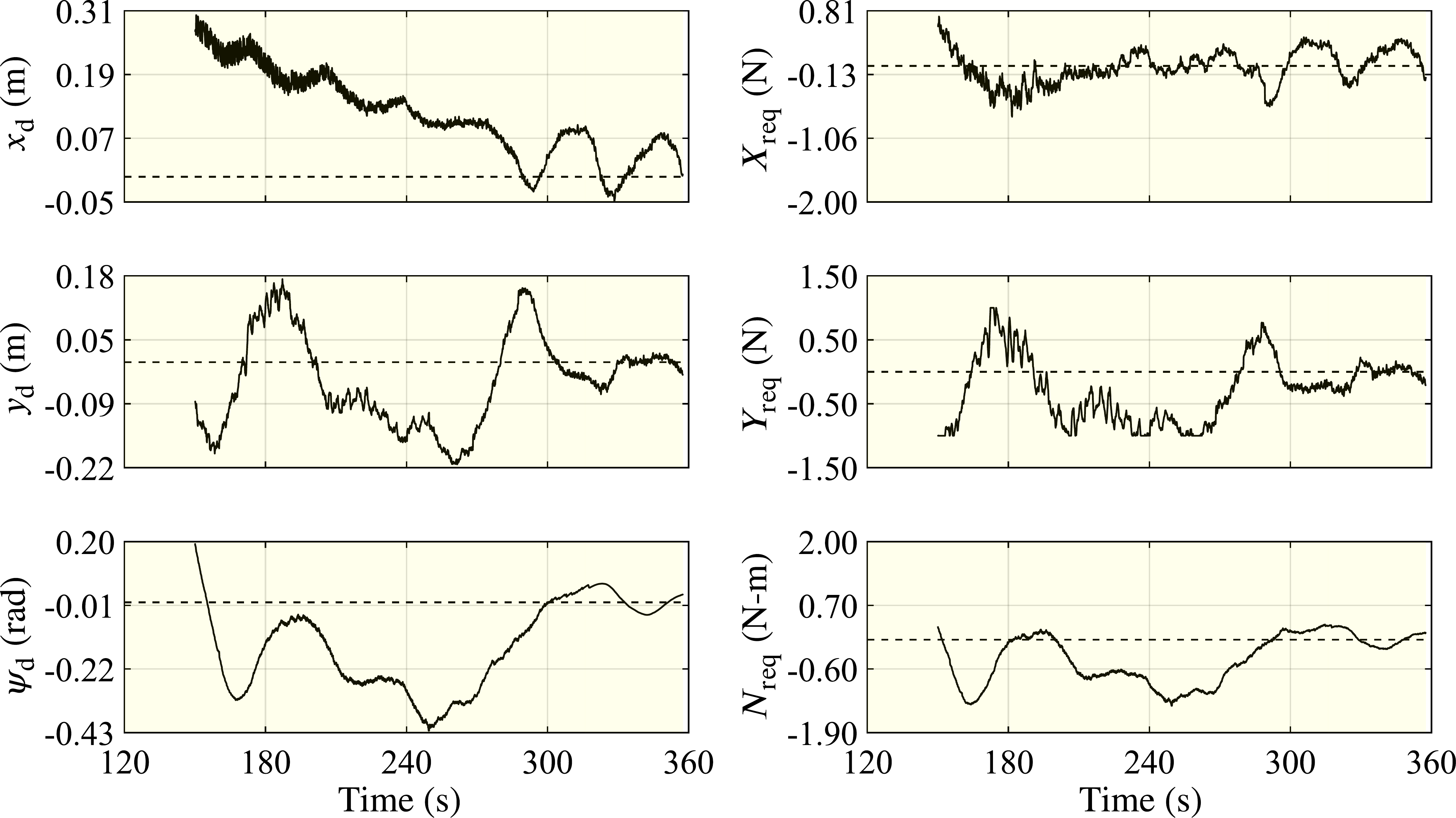}
					\caption{Desired pose in ship-fixed coordinate system and required forces from the VTPS (experiment C).} 
					\label{fig:desiredposeandforces3exp}
				\end{figure}

			    The sole purpose of this experiment is to test the VTPS for a U-turn/180-degree docking. So, only relevant results in the docking stage are shown here. The wind condition during this experiment was calm with 0.1 m/s to 2.0 m/s gusts. As can be seen in Fig. \ref{fig:visual3exp}, the VTPS was able to synthesize the actuator commands such that the ship followed the given U-turn docking poses. In addition, the time histories of the states and the error $\ebold$ are shown in Fig. \ref{fig:states3exp} and Fig. \ref{fig:desiredposeandforces3exp}, respectively.
			    
                When turning, the surge velocity $u$ was kept very low; almost in the same order as the sway velocity $v$ as shown in Fig. \ref{fig:states3exp}. One can observe that the bow thruster was used heavily to rotate the ship while the rudders were used to maintain the low speed. This shows that the current VTPS relies primarily on the bow thruster for an operation that requires a hard turn. Two possible reasons to this are: (a) the linear assumption does not capture the true ability of the VecTwin rudder and (b) the range of the rudder is significantly constrained. It is worth mentioning that the presence of strong gusts could have deteriorated the performance, particularly during the turning at a very low speed. Nevertheless, at the moment, one can consider that the performance of the VTPS is quite promising.

\section{Concluding Remarks}
	\label{Sec:Conclusion}
	This article introduces a novel and experimental low-speed positioning system for a vessel with a VecTwin rudder system and a bow thruster. This system is termed VTPS which stands for VecTwin positioning system. In the VTPS, the propeller revolution is kept constant in forward mode all the time, thus rendering it unavailable as a free actuator from the point of view of control. 
	
	Several scale model experiments were conducted to test the performance of the VTPS directly, thus the term \textit{experimental} to indicate that this system is some kind of a working prototype; in contrast to a numerical simulation. These experiments serve as evidence of the potential of the VTPS. In other words, the hypothesis is directly verified and validated in the experiments rather than going through numerical verification.

    At present, the VTPS utilizes the rudders to transfer the thrust from the main propeller to balance the forces such that the ship hovers without changing the main propeller's revolution number and mode. This is an advantage over the usual DPS arrangement where the auxiliary thrusters actively change their revolution and mode to balance the forces.
    
    The rudders also transfer the thrust to lateral forces and moments to support the bow thruster. However, the linearity assumption of the forces and the limitation on the rudder range results in the VTPS relying on the bow thruster for operations that require a hard turn, e.g., a U-turn docking. This is a disadvantage because the generated lateral forces (and thus the turning moment) are less strong than the usual DPS. This will be discussed and improved in the future.

    Nevertheless, the arrangement: VecTwin rudders, a bow thruster, and a constant single-screw propeller in the current experimental VTPS gave quite satisfactory performances for low-speed operations: path-following and position-keeping, in a calm wind situation with no/weak gusts. With further investigations, it can be a potential and practical addition to the usual DPS. In addition, the experiment results give a direct demonstration of the excellent ability of the VecTwin rudder system to achieve hovering and emergency crash-stopping at a constant forward propeller revolution. Moreover, further technical aspects to improve the performance and reliability of the VTPS will also be covered in future occasions.

\begin{acknowledgements}
This study was conducted as collaborative research with Japan Hamworthy \& Co., Ltd. It was also supported by a Grant-in-Aid for Scientific Research from the Japan Society for the Promotion of Science (JSPS KAKENHI Grant \#19K04858 and \#22H01701).
\end{acknowledgements}

\appendix
\section{Appendix}

\subsection{On the linearity assumption for a wider range of rudder angle: 60 degree to 105 degree}\label{appendixA}
    \begin{figure}[htbp]
        \centering
		\includegraphics[width=1\columnwidth]{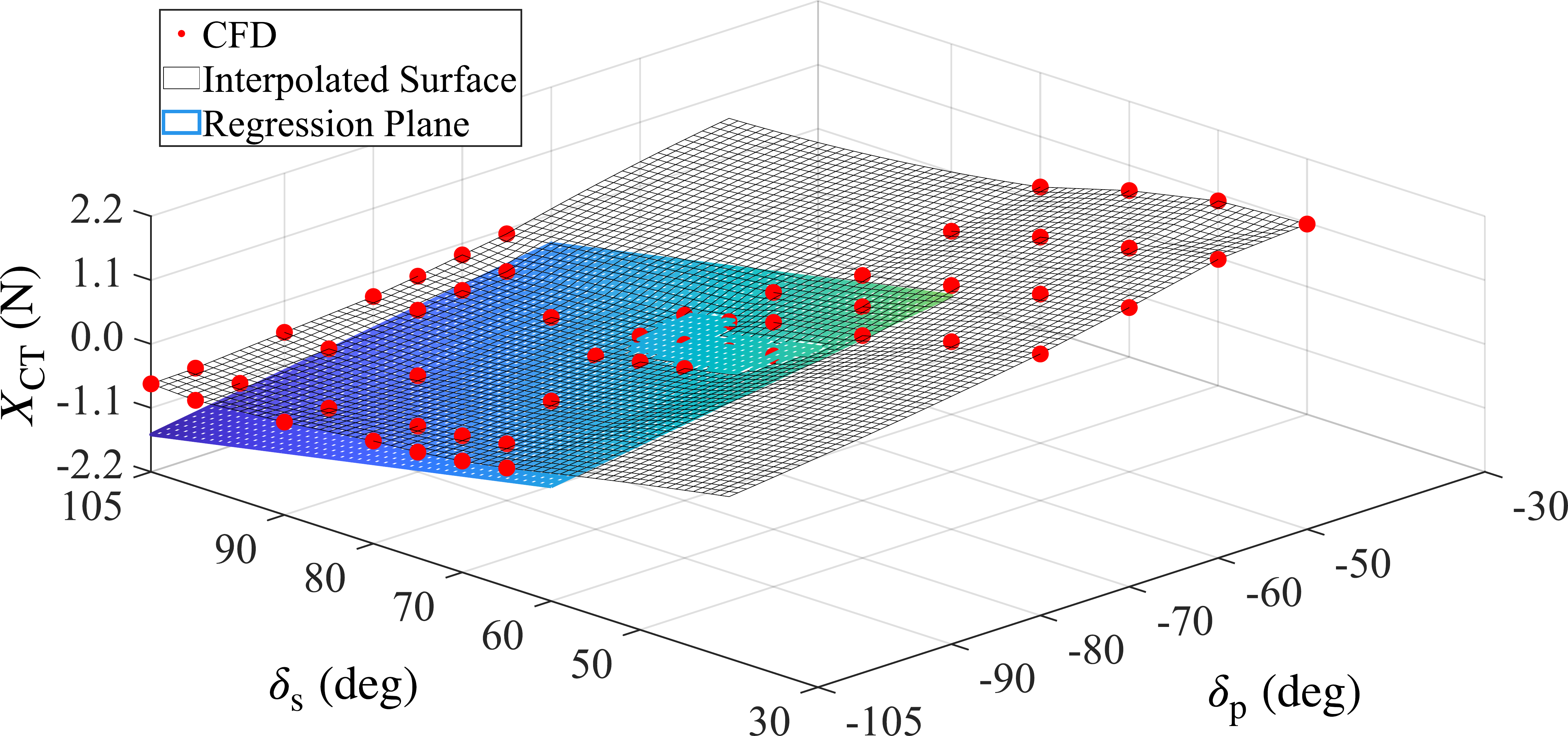}
        \caption{Comparison between the interpolated surface from additional CFD simulations and the regression plane $\lpare{\text{first row of }\Vmtxtilde}$} for rudder angle between 60 degree to 105 degree. 
		\label{fig:appdx1XForce}
	\end{figure}

    \begin{figure}[htbp]
        \centering
		\includegraphics[width=1\columnwidth]{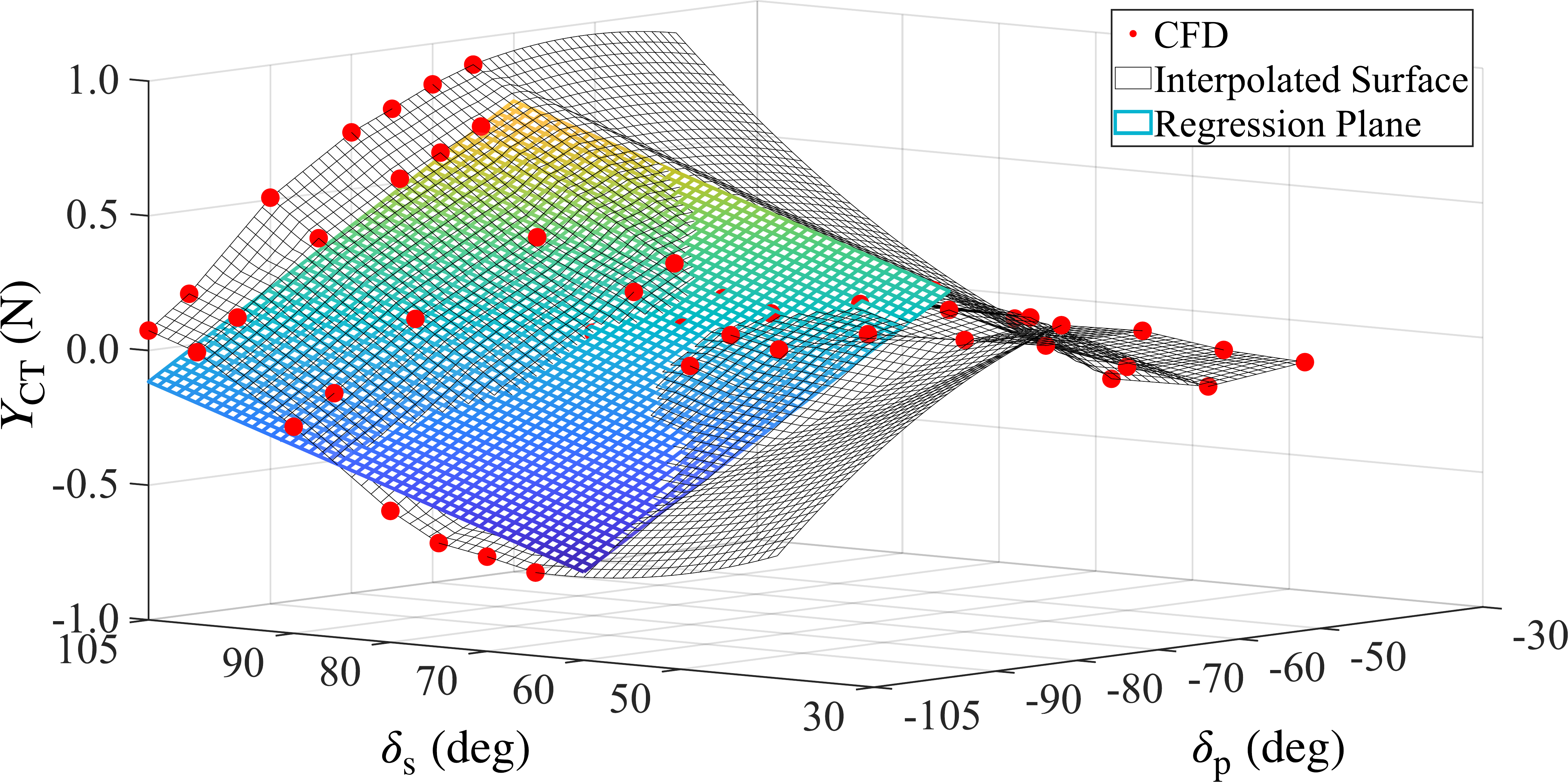}
        \caption{Comparison between the interpolated surface from additional CFD simulations and the regression plane $\lpare{\text{second row of }\Vmtxtilde}$} for rudder angle between 60 degree to 105 degree.  
		\label{fig:appdx1YForce}
	\end{figure}

    The command-force relationship for the rudders $\lpare{\Vmtxtilde}$ was obtained via multiple linear regression of the results from nine CFD simulations (Table \ref{tab:CFDSummary}): pairs of rudder angles $\uboldeta$ (\ref{eq:deltapairvect}) between 70 and 80 degree, i.e., linearization around the hover rudder angle $\uboldth$ (\ref{eq:hoveranglevalue}). This linearity is assumed to be true for a wider range angle: 60 degree to 105 degree. Let $\mathcal{R}$ be the compact set of all $\uboldeta$ within this wider range.
    
    Alternative to physical tests, this assumption is verified by extending the CFD simulations for additional $\uboldeta$: between 40 degree and 105 degree. Let $\mathcal{A}$ be the finite set of this additional $\uboldeta$. One can linearly interpolate the values between two adjacent simulation results (red dots) to form an interpolated surface. This is shown in Fig. \ref{fig:appdx1XForce} and Fig. \ref{fig:appdx1YForce} where the interpolated surface (black) is compared with the regression plane (multicolored): the image of transformation/linear mapping $\Vmtxtilde$ (\ref{eq:vmtxtilde}) given the set $\mathcal{A}$
    as the domain.
    
    From Fig. \ref{fig:appdx1XForce}, the regression plane gives a good estimation of the $\XCT$ for $\uboldeta$ around the hover angle (\ref{eq:hoveranglevalue}): 60 degree to 90 degree. For $\uboldeta$ above 90 degree, the regression overestimates the $\XCT$. This is desirable in low-speed operations because it gives aggressive actions to maintain the low speed. For $\uboldeta$ below 60 degree, as expected, the regression underestimates the $\XCT$, i.e., the slope should be steeper: first row of $\Vmtxtilde$ should be larger in magnitude. This implies that for operations at normal speed and/or smaller $\uboldeta$, a different $\Vmtxtilde$ should be constructed. From Fig. \ref{fig:appdx1YForce}, the regression plane also gives a good estimation of the $\YCT$ for $\uboldeta\in\mathcal{R}$. As one can see, the slopes with respect to $\YCT$ change sign when $\uboldeta$ is lower than 60 degree. Thus, it is safe to say that the linear relationship is acceptable for any $\uboldeta \in \mathcal{R}$: between 60 degree and 105 degree, as verified by the CFD simulations and validated (tested in a closed-loop feedback control scenario) in the experiments.
    
    As a final note, with these additional data points, one may be tempted to obtain a single command-force relationship that covers the whole range of $\uboldeta\in\mathcal{R}$. This is generally not recommended as it will move the intercept of the regression away from the hover rudder angle $\uboldth$, effectively contradicting the fact that at $\uboldth$ the resultant forces are zero that the ship hovers.

\subsection{On the extension of the CFD results as an estimation for any arbitrary ship with VecTwin rudders}
\label{appendixB}
\newcommand{\Arudd}{A_{\mathrm{R}}}
\newcommand{\uR}{u_{\mathrm{R}}}
    \noindent This appendix explains a rather elementary way to extend the CFD results for any arbitrary scale model (or a full-scale ship). Due to the facts that (a) the forces shown in Table \ref{tab:CFDSummary} are due to the propeller-rudder-hull interaction, (b) the propeller revolution $n$ is constant, and (c) the ship's speed is very low, one can nondimensionalize the forces with the rudder properties. It can be done via the following equations (the prime symbol denotes the nondimensionalized quantities),
    \begin{align}\label{eq:nondimension}
        \XCT' = \frac{\XCT}{\rho\Arudd \uR^2} \quad\quad \text{and} \quad\quad 
        \YCT' = \frac{\YCT}{\rho\Arudd \uR^2}, 
    \end{align}
    where $\Arudd$ is the sectional area of the rudder and $\uR$ is the longitudinal fluid inflow velocity at the rudder.
    
    There is still room for discussion on the nondimensionalization of $\XCT$ and $\YCT$ (that include hull forces) with $\Arudd$. At a bollard pull condition, these hull forces can be smaller than those of the rudders and the propeller. Moreover, they depend on the flow reflected by the rudders, thus the nondimensionalization with $\Arudd$. On the other hand, $\uR$ captures the effect of $n$ and can be approximated as \cite{Kang2008},
    \begin{align}
        \uR=k_x\sqrt{\frac{8C_1\mu}{\pi}}nD\Psub,
    \end{align}
    where $k_x$ is the coefficient of fluid inflow acceleration at the rudder, $C_1$ is the intercept of the regression of the propeller thrust coefficient $K_{\mathrm{T}}$, $D\Psub$ is the diameter of the propeller, and $\mu$ is the ratio between $D\Psub$ and the height of the rudder.
    
    Given the geometric properties of the rudder and the propeller, one can extend/estimate the command-force relationship for any arbitrary ship by scaling the $\XCT'$ and $\YCT'$ appropriately following (\ref{eq:nondimension}). Since the relationship is linear, this is equivalent to scaling the $\Vmtxtilde$ (\ref{eq:vmtxtilde}) with the same scaling term. Equivalent dimensional analysis can also be done for the bow thruster, i.e., scaling the $\CB$ (\ref{eq:YB}) according to similarity law. As a caveat, due to the limited studies on this matter, the applicability of this very straightforward dimensional analysis is yet to be validated and hence requires extensive investigations.\\


%
%

\end{document}